\documentclass[prl,noshowpacs,english,superscriptaddress,twocolumn]{revtex4-1}
\usepackage[colorlinks,linkcolor=blue,anchorcolor=blue,citecolor=blue,filecolor=blue,menucolor=blue,runcolor=blue,urlcolor=blue,frenchlinks=blue]{hyperref}
\usepackage{amsmath,amssymb,amsthm,amsfonts,mathrsfs}
\usepackage{graphicx}
\usepackage{pdfpages}
\usepackage{appendix}
\usepackage{lipsum}
\usepackage{slashed}
\usepackage{dcolumn}
\usepackage{mathrsfs}
\usepackage{bm}
\usepackage{color}
\usepackage{lineno,hyperref}
\makeatletter

\newcommand{\Rmnum}[1]{\expandafter\@slowromancap\romannumeral #1@}
\AtBeginDocument{\let\LS@rot\@undefined}
\makeatother
\usepackage{tikz,pgf}
\usetikzlibrary{shadings}
\usepackage{pgfplots,pgfplotstable}

\def \H {\mathcal{H}}

\def \k {\mathbf{k}}

\def \K {\mathcal{K}}

\def \T {\mathcal{T}}
\def \C {\mathcal{C}}

\def \P {\mathcal{P}}

\def \PT {\mathcal{PT}}
\def \CP {\mathcal{CP}}
\def \RT {\mathcal{RT}}

\def \Z {\mathbb{Z}}

\begin{document}
	
	\title{Topological Electromagnetic Effects and Higher Second Chern Numbers in Four- Dimensional Gapped Phases}

	\author{Yan-Qing Zhu}
	\email{yqzhuphy@hku.hk}
	\affiliation{Guangdong-Hong Kong Joint Laboratory of Quantum Matter, Department of Physics, and HKU-UCAS Joint Institute for Theoretical and Computational Physics at Hong Kong,\\ The University of Hong Kong, Pokfulam Road, Hong Kong, China}
	
	
	\author{Zhen Zheng}
	
	\affiliation{Guangdong-Hong Kong Joint Laboratory of Quantum Matter, Frontier Research Institute for Physics, South China Normal University, Guangzhou 510006, China}
	\affiliation{Guangdong Provincial Key Laboratory of Quantum Engineering and Quantum Materials, SPTE, South China Normal University, Guangzhou 510006, China}

	\author{Giandomenico Palumbo}
	\email{giandomenico.palumbo@gmail.com }
	\affiliation{School of Theoretical Physics, Dublin Institute for Advanced Studies,
		10 Burlington Road, Dublin 4, Ireland}
	
	\author{Z. D. Wang}
	\email{zwang@hku.hk}
	\affiliation{Guangdong-Hong Kong Joint Laboratory of Quantum Matter, Department of Physics, and HKU-UCAS Joint Institute for Theoretical and Computational Physics at Hong Kong,\\ The University of Hong Kong, Pokfulam Road, Hong Kong, China}

	\date{\today}
	
	\begin{abstract}
		Higher-dimensional topological phases play a key role in understanding the lower-dimensional topological phases and the related topological responses through a dimensional reduction procedure. In this work, we present a Dirac-type model of four-dimensional (4D) $\Z_2$ topological insulator (TI) protected by $\CP$-symmetry, whose 3D boundary supports an odd number of Dirac cones. A specific perturbation splits each bulk massive Dirac cone into two valleys separated in energy-momentum space with opposite second Chern numbers, in which the 3D boundary modes become a nodal sphere or a Weyl semimetallic phase. By introducing the electromagnetic (EM) and pseudo-EM fields, exotic topological responses of our 4D system are revealed, which are found to be described by the (4+1)D mixed Chern-Simons theories in the low-energy regime. Notably, several topological phase transitions occur from a $\CP$-broken $\Z_2$ TI to a $\Z$ TI  when the bulk gap closes by giving rise to exotic double-nodal-line/nodal-hyper-torus gapless phases. Finally, we propose  to probe experimentally these topological effects in cold atoms.
	\end{abstract}
	
	\maketitle
	
	\emph{{\color{blue}Introduction.}---}
	The prediction and discovery of topological insulators (TIs) and topological semimetals (TSMs) have led to an explosion of activity in  studying  topological aspects of band structures in the past decade. Nowadays, topological phases of matter have been at the forefront of the condensed matter and artificial systems\cite{Hasan,XLQi2011,Armitage,DWZhang2018,Cooper2019,YXu2019,Ozawa2019b,JLiu2020}. One reason for the excitement is that these topological phases beyond Landau's  spontaneous symmetry breaking theory are protected by certain symmetries which supports nontrivial boundary states associated with a topological invariant in the bulk. Another significant feature relies on their corresponding topological electromagnetic (EM) responses which are described by topological field theories. For instance, the $(2+1)$-D Chern-Simons theory describes the quantum anomalous Hall effect in $2$D Chern insulators\cite{Landsteiner2016}, the $(3+1)$D axion field theory describes the magnetoelectric effect in 3D $\Z_2$ TIs\cite{XLQi2008}, and the $(3+1)$D mixed axion theory \cite{Palumbo2014} describes the EM response in certain 3D topological crystalline insulators\cite{Ramamurthy2017}.  Similarly, TSMs also exhibit topological transport phenomena described by the mixed Chern-Simons/axion theory \cite{Ramamurthy2015,Grushin2012,Vazifeh2013,Pikulin2016,Behrends2019} and can be understood via quantum anomalies\cite{Bertlmann}, such as parity anomaly in 2D and 4D TSMs\cite{ZLin2019,YQZhu2020}, chiral anomaly in 3D Weyl semimetals\cite{Zyuzin2012,CXLiu2013}, and $\Z_2$ and chiral anomalies in 3D Dirac semimetals\cite{Burkov2016}.

	On the other hand, higher-dimensional topological phases (HDTPs) are much less explored due to their impossible realization in solid-state materials. However, 4D topological phases can be implemented in synthetic matter as recently shown in Refs\cite{Price2015,Petrides2018,Lee2018,Price2020,WCheng2021,YWang2020,LLi2019}. Moreover, HDTPs play an important theoretical role in lower-dimensional topological phases. For instance, the well-known 2D and 3D TIs can be obtained from a 4D time-reversal-invariant (TRI) insulator through dimensional reduction as well as their effective field-theoretical descriptions\cite{XLQi2008}. The 3D boundaries of such a $\Z$-class 4D TI supports an odd number of Weyl points with same chirality which can not be realized in any 3D systems due to the Nielsen-Ninomiya no-go theorem. Inspired by this idea, one may wonder: \emph{Does there exist a 4D TI phase that supports an odd number of different types of nodal structures on its 3D boundaries? Are there novel topological responses if this system does exist?}
	
\begin{figure}[http]\centering
		\includegraphics[width=8.4cm]{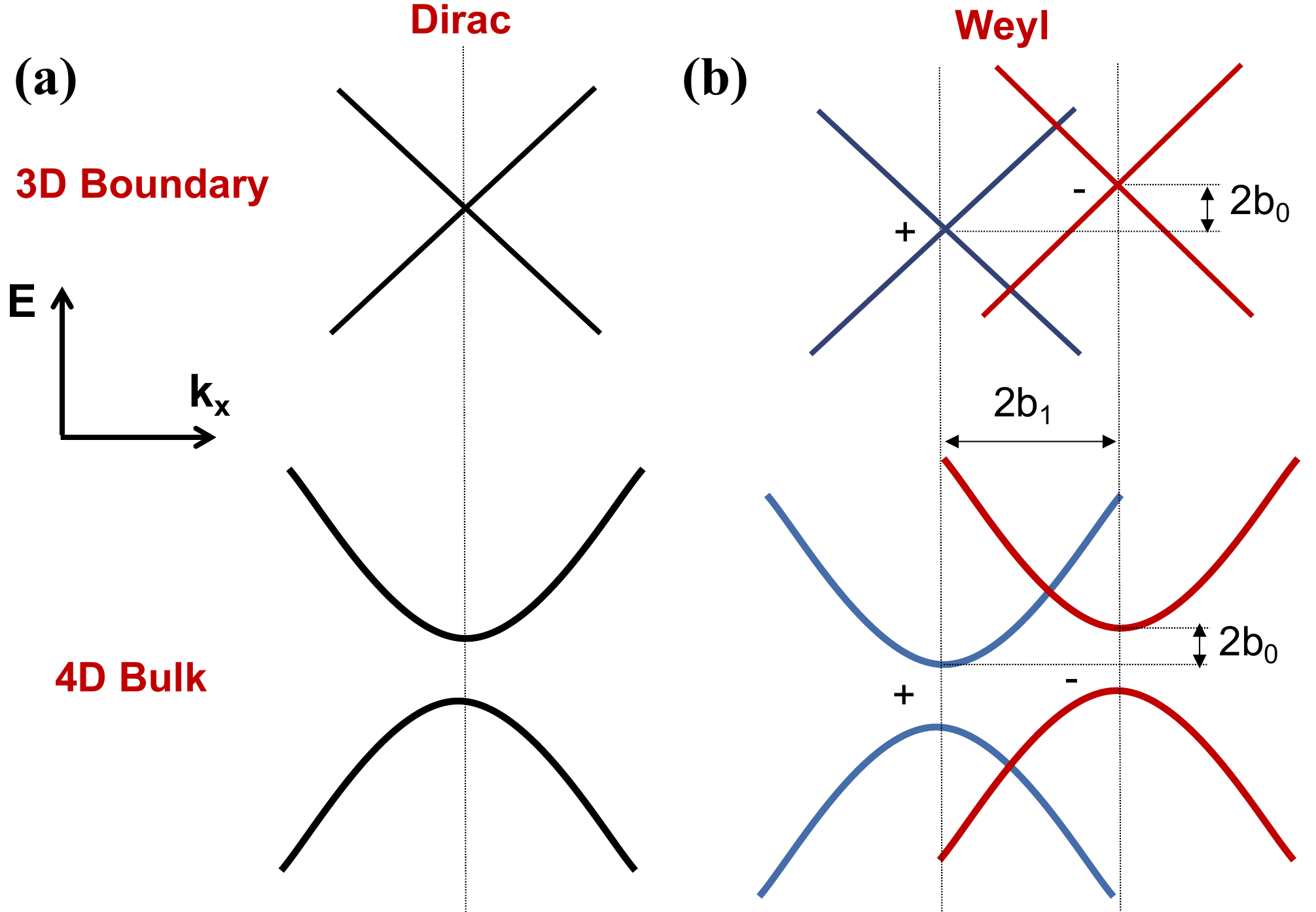}
		\caption{ Schematic of low-energy spectra for $2<m<4$ near $k=0$. (a) (Up panel) A real Dirac cone as a boundary state on the 3D boundary when considering open boundary condition along $w$-direction while a double massive Dirac cone as the bulk spectrum is presented in the bottom panel. (b) Up (Bottom) panel shows that vector field $b_{\mu}$ lifts the degeneracy of the 3D boundary massless (4D bulk massive) Dirac cone which induces two Weyl points (valleys) with opposite chirality (second CN) separated in energy and in momentum  with difference $2b_0$ and $2b_1$, respectively.}
		\label{BDmode}
	\end{figure}

	In this Letter, we answer positively to both questions. We first present a $\Z_2$ 4D TI model that supports an odd number of real Dirac points on its 3D boundary normal to the fourth-dimensions, which are
	protected by  $\CP$-symmetry beyond the ten-fold way classification \cite{Chiu2016}($\C$ and $\P$ denote particle-hole and inversion symmetries, respectively). The bulk $\Z_2$ invariant $\nu_2$ can be defined by the second spin Chern number (CN), i.e., the higher-dimensional generalization of the spin CN in 2D TRI insulators without spin-orbit-coupling (SOC). Subsequently, we will introduce a perturbation which will break the degeneracy of the bulk spectrum forming two valleys with opposite second CNs. In this case $\nu_2$ remains unchanged and is associated with the second valley CN  which is defined as the half difference of the second CNs for two valley indices instead of ``spin". On the 3D boundary, each real Dirac point will split into two Weyl points with opposite chirality in the separation of energy forming a Weyl nodal sphere or of momentum forming a Weyl semimetallic phase, see Fig. \ref{BDmode}. Moreover, we will show that such a $\Z_2$ phase has some novel electromagnetic responses upon applying the EM and pseudo-EM fields\cite{CXLiu2013,Cortijo2015,Shapourian2015,Grushin2016}.  The field theoretical-description of ``\emph{4D quantum spin Hall effect}" in our model is also developed. We identify two novel quantum effects, coined, ``\emph{valley-induced electromagnetic effect}", and ``\emph{4D quantum valley Hall effect}". These bulk responses are consistent with the anomaly equations of the 3D boundary Dirac/Weyl modes arising from the $\Z_2$/chiral anomaly. Furthermore, we explore several types of topological phase transitions from a $\Z_2$ TI to a $\Z$ TI where the phase transition occurs only when a bulk gap closes. We will show that there are many exotic topological phases that appear during the transition processes including the double-nodal-line(DNL)/nodal-hyper-torus(NHT) semimetallic phases and the 4D quantum Hall insulator (QHI) phases with higher second CNs. Finally, we will propose to realize such a 4D model and to detect the predicted responses in a 3D optical lattice with an extra periodic parameter using ultracold atoms.
	
	\emph{{\color{blue}Model and bulk topology.}---}
	Let us  start with the minimal model of a 4D $\Z_2$ TI which takes the form,
	\begin{equation}\label{RCIHam}
		\mathcal{H}_{0}(k)= d_x\Gamma_1+d_y\Gamma_2+d_z\Gamma_3+d_w\Gamma_4+d_m\Gamma_0,
	\end{equation}
	with the Bloch vector being
	$d_i=\sin k_i$ and $d_m=m-\sum_i\cos k_i$
	with $i=x,y,z,w$. The $8\times 8$ matrices $\Gamma_i$ satisfying a Clifford algebra are presented in Ref. \cite{note0}.
	This system hosts two four-fold degenerate bands with the spectrum,
	$E_{\pm}=\pm\sqrt{d_x^2+d_y^2+d_z^2+d_w^2+d_m^2}$.

	
	This model preserves the $\CP$-symmetry, i.e.,
	$\left\{\CP,\mathcal{H}_{0}\right\}=0$,
	with $\CP=iG_{112}\mathcal{K}$ satisfying $(\CP)^2=-1$.  We label $G_{ijk}=\sigma_{i}\otimes\sigma_j\otimes\sigma_k$ hereafter and $\mathcal{K}$ as the complex conjugate operator. Thus this system is characterized by a $\mathbb{Z}_2$ invariant\cite{YXZhao2016}.
	We next move to discuss the bulk topology of this system.
	The Hamiltonian commutes with a matrix $\Gamma_5=G_{332}$, i.e.,
	$\left[\Gamma_5,\mathcal{H}_{0}\right]=0$.
	It implies that this model can be block diagonalized after rotating $\Gamma_5$ into $G_{300}$ through a unitary matrix $U=\exp\left(-\frac{i\pi}{4}G_{100}\right)\exp\left(\frac{i\pi}{4}G_{132}\right)$, and is written in the form of
	$\mathcal{H}_{BD}=U\mathcal{H}_{0}U^{-1}=\H_+\oplus\H_-$,
	where each block Hamiltonian is given by
	
		\begin{equation}\label{BlockHam}
			\mathcal{H}_{\pm}(k)=-d_xG_{33}+ d_yG_{10}-d_zG_{31}-d_wG_{20}\pm d_mG_{32},
	\end{equation}
	with $G_{ij}=\sigma_i\otimes\sigma_j$.
	
	$\mathcal{H}_{BD}$ preserves $\CP$-symmetry with
	{$\CP=iG_{120}\mathcal{K}$}.
	Although $\mathcal{H}_{\pm}$ breaks $\CP$-symmetry, each block Hamiltonian preserves the time-reversal-symmetry (TRS) and thus falls into class AII with {$\T=iG_{20}\K$} satisfying $\T^2=-1$. Therefore, each block model is a 4D QHI characterized by the second CN\cite{XLQi2008},
	\begin{equation}
		\begin{split}
			C_2^{\pm}&=\frac{1}{32\pi^2}\int_{\mathbb{T}^4}d^4k~\epsilon^{\mu\nu\rho\sigma} \text{tr}\left(\mathcal{F}^{\pm}_{\mu\nu}\mathcal{F}^{\pm}_{\rho\sigma}\right),\\
		\end{split}
	\end{equation}
	with the values $C_2^{\pm}=\pm 3\text{sgn}(m)$ for  $0<|m|<2$, $C_2^{\pm}=\mp \text{sgn}(m)$ for $2<|m|<4$, and $C_2^{\pm}=0$ elsewhere.
	Here 
	the non-Abelian Berry curvature
	$\mathcal{F}^{\pm}_{\mu\nu}=\partial_{\mu}\mathcal{A}^{\pm}_{\nu}-\partial_{\nu}\mathcal{A}^{\pm}_{\mu}-i\left[\mathcal{A}^{\pm}_{\mu},\mathcal{A}^{\pm}_{\nu}\right]$,
	where $(\mathcal{A}^{\pm}_{\mu})^{\alpha\beta}=i\langle u^{\pm}_{\alpha}|\partial_{\mu}|u^{\pm}_{\beta}\rangle$ denotes the non-Abelian Berry connection defined by the occupied eigenstates $|u^{\pm}_{\alpha}\rangle$ of each block (we set the Fermi energy at $\epsilon_F=0$).
	Therefore, although the total second CN is zero namely $C_2=C_2^++C_2^-=0$, we can define a $\mathbb{Z}_2$ number given by
	\begin{equation}\label{Z2inv}
		\nu_2=C_{2s} ~\text{mod}~2,
	\end{equation}
	with the second spin CN\cite{Petrides2022} being $C_{2s}=(C_2^+-C^-_2)/2$ which can be regarded as a generalization of the first spin CN in a 2D TRI system without SOC. Note that one can also define the $\mathbb{Z}_2$ number $\nu_2$ in terms of the Green function for the original model $\mathcal{H}_0$\cite{note4}.
	
	This system enjoys more extra symmetries\cite{SM}, e.g., $\RT$, mirror, time-reversal, particle-hole, and chiral symmetries; see Supplemental Material (SM) for details. Thus it implies that this model will host a very rich phase diagram when we introduce some extra symmetry-protected/broken perturbations.
	In the following part, we mainly focus on the $\mathbb{Z}_2$ topological phases and discuss the perturbations that commuting with $\Gamma_5$
	which means these terms can be block diagonalized, i.e., {\color{red}$\nu_2$} is still well-defined and unchanged unless there is a gap closing phase transition in the bulk. For concreteness, we introduce the $\CP$-broken but $\Gamma_5$-protected perturbation,
	\begin{equation}
		\begin{split}
			\Delta&=\Gamma_5(b_{0}+\sum_{i=1}^3b_i\Gamma_i).
		\end{split}
	\end{equation}

	
	\emph{{\color{blue}Boundary Physics.}---}Without loss of generality, we mainly address the 3D boundary physics by considering a boundary perpendicular to $w$-direction with a lattice length $L_w$.  The cases for open boundary conditions along the other directions are presented in the SM \cite{SM}.
	Now let us  first focus on the case of $\mathcal{H}_{0}$ where $2<m<4$ and thus $C_2^{\pm}=\mp1$.
	The effective boundary Hamiltonian at $L=0$ is given by
	$\mathcal{H}_{eff}=-\sin k_xG_{13}+\sin k_yG_{30}-\sin k_zG_{11}$.
	It supports a real Dirac point at the origin with the expanded Hamiltonian,
	\begin{equation}
		\mathcal{H}_{RD}(\k)=-k_xG_{13}+k_yG_{30}-k_zG_{11}.
	\end{equation}
	Note that this model preserves $\PT$-symmetry, i.e.,
	$\left[\PT,\mathcal{H}_{RD}\right]=0$, with $\PT=\mathcal{K}$ satsfying $(\PT)^{2}=+1$.
	Thus this model presents a real Dirac monopole with a  $\mathbb{Z}_2$ classification and carries a monopole charge $\nu_R=1$ defined in terms of the real CN \cite{YXZhao2017} or the first Euler number \cite{Ahn2018,Bouhon2020a,Bouhon2020b}. Moreover, one can see that a real Dirac point consists of two Weyl points with opposite chirality under the rotation representation\cite{note3}.

	Next we add the above-mentioned term $\Delta$ to $\H_0$, i.e.,
	$\mathcal{H}_1=\mathcal{H}_0+\Delta$.
	The system now hosts two Dirac valleys with energy and momentum differences $\delta E=2b_0$ and $\delta k=2\mathbf{b}$ in the bulk, respectively, see Fig. \ref{BDmode}.
	In the block diagonal representation, each block Hamiltonian still hosts the unchanged second CN $C_2^{\pm}$ and thus we have $\nu_2=C_{2v}~\text{mod}~2$ with $C_{2v}=C_{2s}$ which is nothing but  the second valley CN.
	To be more clear, we first discuss the case when only $b_0$ is non-zero.
	Now the total low-energy 3D boundary Hamiltonian becomes
	\begin{equation}\label{NSHam}
		\mathcal{H}_{NS}(\k)=\mathcal{H}_{RD}(\k)+b_0\gamma_5,
	\end{equation}
	with $\gamma_5=G_{32}$. Note that $\gamma_5$ commutes with $\mathcal{H}_{RD}$, i.e, $\left[\gamma_5,\mathcal{H}_{RD}\right]=0$. This model represents a 3D Weyl nodal sphere with spectrum
	$E_{\pm}=\pm|\k|\pm|b_0|$,
	exhibiting a band degeneracy at the Fermi level $E=0$
	on a sphere defined by $|\k|=|b_0|$ with $|\k|=\sqrt{k_x^2+k_y^2+k_z^2}$.
	This model breaks $\PT$ but keeps the $\gamma_5$-symmetry, and thus the nodal sphere carries a $\mathbb{Z}$ monopole charge\cite{Turker2018,Salerno2020}.

	In the case when only $\mathbf{b}=(b_1,b_2,b_3)$ is nonzero, we obtain a Weyl semimetallic phase on its 3D boundary with the low-energy effective Hamiltonian given by
	\begin{equation}
		\mathcal{H}_{WS}=\H_{RD}(\k)+b_1G_{21}+b_2G_{02}+b_3G_{23}.
	\end{equation}
	Its spectrum reads
	$E_{\pm}=\pm\sqrt{(k_x\pm b_1)^2+(k_y\pm b_2)^2+(k_z\pm b_3)^2}$,
	which represents a pair of Weyl points with opposite charge/chirality at $\k_W^{\pm}=\pm(b_1,b_2,b_3)$.
	
	In other words, the 3D boundary physics of $\H_1$ describes a pair of Weyl points with opposite chirality separated in energy-momentum space with $\delta E=2b_0$ and $\delta \k=2\mathbf{b}$, as shown in Fig.\ref{BDmode}.
	Generally, there are $|C_{2v}|$ gapless boundary modes on its 3D boundary when $m$ varies in different regions~\cite{note6}.
	Notice that the 2D (0D) topological charge implies that the boundary band structure with an odd number of real Dirac points (nodal spheres) cannot be realized by any 3D systems, and it can only exists at the boundary of a 4D gapped system.

	\emph{{\color{blue}Topological responses.}---}
	For later convenience, we here consider the simplest case in the block diagonal representation, i.e., $\mathcal{H}_{BD}$.
	To calculate the continuum response we couple each continuum Hamiltonian with its own gauge field $A_{\mu}^{(a)}$
	via $k_{\mu}\rightarrow k_{\mu}+A^{(a)}_{\mu}$.
	The topological action of the 4D QHI in each block is described by the (4+1)D Chern-Simons theory \cite{XLQi2008}, i.e.,
	\begin{equation}
		S^{(a)}_{\text{eff}}=\frac{C^{(a)}_2}{24\pi^2}\int~d^5k ~\epsilon^{\mu\nu\lambda\rho\sigma}A^{(a)}_{\mu}\partial_{\nu}A^{(a)}_{\lambda}\partial_{\rho}A^{(a)}_{\sigma},
	\end{equation}
	where $a=\pm$ for each block, and
	$A_{\mu}=\frac{1}{2}\left[A^{(+)}_{\mu}+A^{(-)}_{\mu}\right]$,
	$\tilde{A}_{\mu}=\frac{1}{2}\left[A^{(+)}_{\mu}-A^{(-)}_{\mu}\right]$.
	The symmetric combination of the gauge fields $A^{(+)}_{\mu}$ and $A^{(-)}_{\mu}$ represents the
	usual EM field $A_{\mu}$; while the antisymmetric combination
	is a spin gauge field $\tilde{A}_{\mu}$,
	such that we obtain a new kind of mixed Chern-Simons action,
	\begin{equation}
		\begin{aligned}\label{MixedCS1}
			S_{\text{eff}}
			&=\frac{C_{2s}}{4\pi^2}\int d^5x
			\epsilon^{\mu\nu\lambda\rho\sigma}\tilde{A}_{\mu}\partial_{\nu}A_{\lambda}\partial_{\rho}A_{\sigma}\\
			&+\frac{C_{2s}}{12\pi^2}\int d^5x
			\epsilon^{\mu\nu\lambda\rho\sigma}\tilde{A}_{\mu}\partial_{\nu}\tilde{A}_{\lambda}\partial_{\rho}\tilde{A}_{\sigma}.
		\end{aligned}
	\end{equation}
	This action can also be derived in a direct diagrammatic calculation by evaluating the diagram \cite{YQZhu2020,note1}. Varying it with respect to $A_{\mu}$ and $\tilde{A}_{\mu}$, respectively, we obtain the  corresponding charge and spin currents, i.e., $J^{\mu}=\delta S_{\text{eff}}/\delta A_{\mu}$, and $\tilde{J}^{\mu}=\delta S_{\text{eff}}/\delta \tilde{A}_{\mu}$.
	For instance, considering a simple field configuration,
	\begin{equation}\label{gauge}
		\tilde{A}_{\mu}=0,~
		A_{\mu}=(-zE_z,-yB_z,0,0,0),
	\end{equation}
	we obtain,
	\begin{equation}\label{4DQSHE}
		J^w=0,~
		\tilde{J}^{w}=\frac{C_{2s}}{4\pi^2}
		E_zB_z.
	\end{equation}
	Here $\mathbf{E}$ and $\mathbf{B}$ denote the components of the usual EM field defined later.
	$\tilde{J}^w$ denotes the spin current upon applying an EM field which can be regarded as ``\emph{4D quantum spin Hall effect}". Note that the index ``spin" above may also include but not limited to the following internal degree of freedom of different physical objects, e.g.,  hyper-states and orbits of atoms, etc.
	
	The above results can actually be understood by connecting to the 2D cases \cite{SM}. Integrating over $x,y$ dimensions with periodic boundary conditions and assuming $E_z$ does not
	depend on $(x,y)$, we have
	\begin{equation}
		\begin{split}\label{2Dspin}
			\int dxdy\tilde{J}^w= \frac{C_{2s}}{2\pi}N_{xy}E_z,\\
		\end{split}
	\end{equation}
	where $N_{xy}=\int dxdyB_z/2\pi$ denotes the number of flux quanta
	through the $xy$ plane, which is always quantized to be an
	integer.  Thus, we can understand this result in a
	$\mathbb{Z}_2$ (4+1)D insulator with the second spin CN $C_{2s}$. This formula denotes the quantum spin Hall effect  with a  spin Hall conductance $C_{2s} N_{xy} /2\pi$ in the $zw$ plane induced by a magnetic field with flux $2\pi N_{xy}$ in the normal
	$xy$ plane.
	
	On the other hand, since this model supports $|C_{2s}|$ Dirac cones on its 3D boundary, we can explore the 3D boundary response properties from the bulk. Integrating out with respect to $w$ (picking the gauge as Eq. \eqref{gauge}) in $S_{\text{eff}}$, we obtain the boundary term
	\begin{equation}\label{bdaction}
		S_{BD}=\frac{C_{2s}}{12\pi^2}\int d^4x\epsilon^{\mu\nu\rho\sigma}\left(3\tilde{A}_{\mu}A_{\nu}\partial_{\rho}A_{\sigma}+\tilde{A}_{\mu}\tilde{A}_{\nu}\partial_{\rho}\tilde{A}_{\sigma}\right),
	\end{equation}
	where $\mu=t,x,y,z$.  The corresponding  currents are given by
	\begin{equation}
		\begin{split}\label{bdcurrent}
			j^z=0,~\tilde{j}^{z}=\frac{ C_{2s}}{4\pi^2}zE_zB_z.
		\end{split}
	\end{equation}
	After varying above currents with respect to $z$, we obtain  $\partial_{z}j^z=J^w$ and $\partial_{z}\tilde{j}^z=\tilde{J}^w$ are the same as those presented in Eq. \eqref{4DQSHE}. This result can be also directly derived from the boundary Dirac Hamiltonian\cite{SM}. We emphasize that such a non-zero spin current $\tilde{j}^z$ only appears on the boundary of 4D $\mathbb{Z}_2$  TIs which stems from the odd number of (real) Dirac cone structure that cannot be realized by any 3D systems\cite{YXZhao2017}


	In what follows we consider the case of $\mathcal{H}_{1}$.
	By coupling each block Hamiltonian with  $A_{\mu}^{(\pm)}$ via $k_{\mu}\rightarrow k_{\mu}+A_{\mu}^{(\pm)}$ and treating $b_{\mu}$ as an axial gauge field with $A_{\mu}^{(\pm)}=A_{\mu}\pm b_{\mu}$, where $b_{\mu}=(b_0,\mathbf{b},0)$,  we have $b_{\mu}\equiv A_{\mu}^5=\frac{1}{2}\left[A_{\mu}^{(+)}-A_{\mu}^{(-)}\right]$.  Consequently, we obtain the effective action just by replacing $\tilde{A}_{\mu}$ with $b_{\mu}$, i.e.,
	\begin{equation}
		\begin{split}
			S_{\text{v,eff}}=\frac{C_{2v}\epsilon^{\mu\nu\lambda\rho\sigma}}{12\pi^2}\int d^5x
			\left(3b_{\mu}\partial_{\nu}A_{\lambda}\partial_{\rho}A_{\sigma}+b_{\mu}\partial_{\nu}b_{\lambda}\partial_{\rho}b_{\sigma}\right).
		\end{split}
	\end{equation}
	Note that this bulk response matches with the EM response of a 4D topological semimetal that hosts two 4D monopoles separated in energy-momentum space where we find precisely half of coefficient occurs\cite{YQZhu2020,note1}.
	The corresponding charge and valley currents are derived from $J^{\mu}=\delta S_{\text{v,eff}}/\delta A_{\mu}$ and $J^{\mu}_5=\delta S_{\text{v,eff}}/\delta b_{\mu}$, respectively.
	Without loss of generality, we take $b_{\mu}=(b_0,b_1(y),b_2,b_3(t),0)$ and $A_{\mu}$ the same as in Eq. \eqref{gauge}. Defining the (pseudo-) magnetic  and (pseudo-) electric fields as
	$\mathbf{B}=\nabla\times\mathbf{A}$ ($\mathbf{B}^5=\nabla\times\mathbf{b}$), and $\mathbf{E}=\partial_t\mathbf{A}-\nabla A_0$ ($\mathbf{E}^5=\partial_t\mathbf{b}-\nabla b_0$), respectively.
	We obtain the charge and valley currents,
	\begin{subequations}\label{bcurrent}
		\begin{align}
			J^{w}&=\frac{C_{2v}}{2\pi^2}(E^5_zB_z+E_zB^5_z),\label{Za}\\
			J^{w}_5&=\frac{C_{2v}}{4\pi^2}\left(E_zB_z+E^5_zB^5_z\right)\label{Zb}.
		\end{align}
	\end{subequations}
	Since the charge current $J^w$ induced by a varying $b_{\mu}$ and magnetic (electric) field, we name it ``\emph{valley-induced magnetic (electric) effect}".  $J_5^w$ is related to the ``\emph{4D quantum valley Hall effect}"  where two valley Hall currents propagate along opposite directions.
	
	We can also integrate $J^w$ and $J_5^w$ with respect to $x,y$ and obtain
	\begin{subequations}
		\begin{align}
			\int dxdyJ^w=\frac{ C_{2v}}{\pi}\left(N_{xy}E^5_z+N^5_{xy}E_z\right),\label{Wa}\\
			\int dxdy J^w_5= \frac{C_{2v}}{2\pi}\left(N_{xy}E_z+N^5_{xy}E^5_z\right),\label{Wb}
		\end{align}
	\end{subequations}
	where $N_{xy}^5=\int dxdyB_z^5/2\pi$ now denotes the number of pseudo-flux quanta
	through the $xy$ plane, which can be also quantized to be an integer as $N_{xy}$ we mentioned above. The understanding of these terms will be similar to the case in Eq. \eqref{2Dspin}. For instance, the second (first) term in Eq. \eqref{Wa} could be treated as the (pseudo-) Hall effect with a quantized  hall conductance $C_{2v} N^5_{xy}(N_{xy}) /\pi$ in the $zw$ plane induced by a pseudo-(usual) magnetic field with flux $2\pi N^5_{xy}(2\pi N_{xy})$ in the normal $xy$ plane. Similarly, the (second) first  term in Eq. \eqref{Wb} represents the (pseudo-) valley Hall effect in the $zw$ plane induced by a (pseudo-) usual magnetic field in the $xy$ plane. 

	From the viewpoint of 3D boundary physics, $b_{\mu}$ splits a Dirac point into two Weyl points with opposite chirality separated in energy and in momentum, i.e., there are $|C_{2v}|$ pairs of Weyl points on the boundary. One can straightforwardly obtain the boundary effective action and the corresponding response currents from the bulk as we did above.  It is possible to show that the 3D boundary supports topological responses upon applying EM and pseudo-EM fields are the same as those ones derived from the 3D Weyl Hamiltonian which has been widely studied in the previous work\cite{Ilan2020}. The boundary response is actually a signature of a bulk response. Interestingly,  when $b_{\mu}$ is constant, we obtain non-zero chiral magnetic and anomalous Hall current in the 3D boundary even though the 4D bulk charge current is zero.

	\emph{{\color{blue}Topological phase transitions.}---}In the presence of $\Delta$, the system $\H_1$ breaks all the symmetries and falls into class A.  Even though each block subsystem breaks TRS, each of them is still a 4D QHI in class A which is quite robust and is characterized by the unchanged second CN $C_2^{\pm}$ and hosts nontrivial boundary modes. So the system continues to host a $\Z_2$ number until a topological phase transition occurs when the bulk gap closes with a critical value $b_{\mu}$.   This indicates a phase transition from a $\Z_2$ TI to a trivial insulator where the corresponding topological response predicted from Eq. \eqref{bcurrent} also changes from nontrivial to trivial.  Moreover, we can add a term $\Delta_a=cG_{302}+b_0G_{332}$ to $\H_0$. The model now falls into class A and goes through a phase transition from a $\Z_2$ TI to a nontrivial DNL/NHT semimetallic phase and then finally becomes a trivial insulator by increasing $|c|$. Such a DNL structure is characterized by the first CN while a NHT is the 3D torus protected by the $\gamma_5$-symmetry associated with a 0D $\Z$-value number\cite{SM}. Furthermore, the term $\Delta_z=\delta_z G_{032}$ that will induce a phase transition from $\Z_2$ to $\Z$ TIs  falls into class AII when $|\delta_z|$ is large enough. This term induces a very rich phase diagram that supports higher second CNs (e.g., $C_2=\pm2,\pm4,\pm6$) by varying $m$ and $\delta_z$\cite{SM}.

	\emph{{\color{blue}Conclusion and outlook.}---}
	We have proposed a novel $\Z_2$ TI model characterized by a $\Z_2$ number $\nu_2$ associated with the second spin (valley) CN, whose 3D boundaries support an odd number (pairs) of real Dirac (Weyl) points, and investigated its topology. In particular, we have revealed several new types of topological responses upon applying the EM and the pseudo-EM fields. Several external terms induce topological phase transitions and give rise to very rich phase diagrams. These topological quantum effects can also appear in the 4D $\Z_2$ TIs in class CII and C~\cite{SM}.
	Note that the predicted responses could be experimental studied in 3D quantum-engineered setups extended by a synthetic/artifical dimension \cite{Ozawa2019}, as could be realized in the photonic/phononic crystals\cite{Ozawa2019b,JLiu2020} where the pseudo-EM field can be induced by a strain field\cite{Rechtsman2013,ZYang2017}, in electric circuits\cite{YWang2020,LLi2019}, or in optical lattices with cold atoms\cite{DWZhang2018,Cooper2019,ZZheng2019,Jamotte2022}, etc. In the SM, we present an experimentally feasible proposal for realizing the $\Z_2$ TI model and detecting the predicted responses in a 3D optical lattice with an external periodic parameter using ultracold atoms~\cite{SM}.
	These results may pave the way for exploring topological responses in the HDTPs  and in artificial systems.
	
	Finally, we note that topological crystalline \cite{LFu2011,Slager2013} or higher-order topological phases \cite{Wieder2020,Schindler2018} can be induced from $\H_0$ and the corresponding effective field-theoretical descriptions are worth to be explored.   By using  the dimensional reduction method, one can explore the 2D $\Z_2$ pumping of $\H_0$ where $\nu_2$ can be measured through the drift of the center of mass of atom clouds\cite{Lohse2018}.  Besides, the investigation of a non-Hermitian 4D $\Z_2$ TI and its topological field theory \cite{Kawabata2021,Sayyad2021} is also one of the possible directions.  Moreover, some interesting physics in spin and anomalous planar Hall systems \cite{Bhlla2021,Cullen2021} can be generalized into 4D while our model with interacting may reveal more novel physics regarding on the current progress\cite{Marzuoli2012,Gian2013,Gian2014, Rylands2021,Cirio2014,Rylands2021}. 

	\begin{acknowledgements}
		This work was supported by the Key-Area Research and Development Program of Guangdong Province (Grant No. 2019B030330001), NSFC/RGC JRS grant (N-HKU774/21), and the CRF of Hong Kong (C6009-20G).
	\end{acknowledgements}

\widetext
\clearpage


\section*{Supplementary material}

# Supplemental material for "Topological Electromagnetic Effects and Higher Second Chern Numbers in Four-Dimensional Gapped Phases"

Yan-Qing Zhu,[1, *] Zhen Zheng,[2, 3] Giandomenico Palumbo,[4, †] and Z. D. Wang[1, ‡]

[1]*Guangdong-Hong Kong Joint Laboratory of Quantum Matter, Department of Physics,
and HKU-UCAS Joint Institute for Theoretical and Computational Physics at Hong Kong,
The University of Hong Kong, Pokfulam Road, Hong Kong, China*
[2]*Guangdong-Hong Kong Joint Laboratory of Quantum Matter,
Frontier Research Institute for Physics, South China Normal University, Guangzhou 510006, China*
[3]*Guangdong Provincial Key Laboratory of Quantum Engineering and Quantum Materials,
SPTE, South China Normal University, Guangzhou 510006, China*
[4]*School of Theoretical Physics, Dublin Institute for Advanced Studies, 10 Burlington Road, Dublin 4, Ireland*
(Dated: October 10, 2022)

---

* yqzhuphy@hku.hk
† giandomenico.palumbo@gmail.com
‡ zwang@hku.hk

# CONTENTS

## S-I. EXTRA SYMMETRIES

The model preserves an extra symmetry, say, $\mathcal{RT}$-symmetry[1], which acts on the Hamiltonian as,

$$\mathcal{RT}\mathcal{H}_0(\mathbf{k}, k_w)(\mathcal{RT})^{-1} = \mathcal{H}_0(\mathbf{k}, -k_w), \tag{S1}$$

where $\mathcal{RT} = \mathcal{K}$, and $(\mathcal{RT})^2 = +1$. In other words, this model preserves mirror symmetry $\mathcal{M}_w$, i.e.,

$$\mathcal{M}_w \mathcal{H}(\mathbf{k}, k_w)\mathcal{M}_w^{-1} = \mathcal{H}(\mathbf{k}, -k_w), \tag{S2}$$

with $\mathcal{M}_w = iG_{312}$ satisfying $(\mathcal{M}_w)^2 = -1$. At the mirror-invariant plane $k_w = 0, \pi$, the 3D subsystems are the 3D real/Takagi topological insulators with $\mathcal{PT} = \mathcal{K}$[2, 3].

Besides, one can check that $\mathcal{H}_0$ also preserves the time-reversal symmetry (TRS), particle-hole symmetry (PHS), and chiral symmetry (CS) with the operators

$$\begin{aligned} \mathcal{T} &= G_{012}\mathcal{K}, \ \ \mathcal{T}^2 = -1, \\ \mathcal{C} &= G_{212}\mathcal{K}, \ \ \mathcal{C}^2 = +1, \\ \mathcal{S} &= G_{200}, \ \ \ \ \mathcal{S}^2 = +1, \end{aligned} \tag{S3}$$

respectively. Thus this model falls into class DIII and seems to be trivial according to the ten-fold way classifications[4]. However, this model hosts a $\mathbb{Z}_2$ invariant since it preserves the $\mathcal{CP}$-symmetry as mentioned in the main text.

## S-II. BOUNDARY PHYSICS

We consider an open boundary condition along $w$-direction, and first apply the inverse Fourier transform for $k_w$ to get the first quantized Hamiltonian with the $w$-dimension in real space,

$$\mathcal{H} = \sum_{i=1}^{3} \sin k_i \Gamma_i + \frac{1}{2i}(S_w - S_w^\dagger)\Gamma_4 + \left[ m - \sum_{i=1}^{3} \cos k_i - \frac{1}{2}(S_w + S_w^\dagger) \right]\Gamma_0 \tag{S4}$$

where $S_w$ is the translation operator along the $w$-direction, and $S_w^\dagger$ is that along the $-w$-direction. So, $S|i\rangle = |i+1\rangle$ and $S^\dagger|i\rangle = |i-1\rangle$ with integer $i$ numbering lattice site of the $w$-dimension, and accordingly the matrices are

$$S_w = \begin{pmatrix} \ddots & \vdots & \vdots & \vdots & \vdots & \cdots \\ \ddots & 0 & 0 & 0 & 0 & \cdots \\ \cdots & 1 & 0 & 0 & 0 & \cdots \\ \cdots & 0 & 1 & 0 & 0 & \cdots \\ \cdots & 0 & 0 & 1 & 0 & \cdots \\ \cdots & \vdots & \vdots & \vdots & \vdots & \ddots \end{pmatrix}, \quad S_w^\dagger = \begin{pmatrix} \ddots & \ddots & \vdots & \vdots & \vdots & \cdots \\ \cdots & 0 & 1 & 0 & 0 & \cdots \\ \cdots & 0 & 0 & 1 & 0 & \cdots \\ \cdots & 0 & 0 & 0 & 1 & \cdots \\ \cdots & 0 & 0 & 0 & 0 & \ddots \\ \cdots & \vdots & \vdots & \vdots & \vdots & \ddots \end{pmatrix}. \tag{S5}$$

If the boundary is opened at $w = 0$ and the system being on the positive part of the $w$-axis ($w > 0$), the corresponding Hamiltonian is given by

$$\widehat{\mathcal{H}} = \sum_{i=1}^{3} \sin k_i \Gamma_i + \frac{1}{2i}(\widehat{S}_w - \widehat{S}_w^\dagger)\Gamma_4 + \left[ m - \sum_{i=1}^{3} \cos k_i - \frac{1}{2}(\widehat{S}_w + \widehat{S}_w^\dagger) \right]\Gamma_0, \tag{S6}$$

where still $\widehat{S}_w|i\rangle = |i+1\rangle$ with $i = 0, 1, 2, \cdots$, but $\widehat{S}_w^\dagger|0\rangle = 0$ and $\widehat{S}_w^\dagger|i\rangle = |i-1\rangle$ with $i \geq 1$. Explicitly,

$$\widehat{S}_w = \begin{pmatrix} 0 & 0 & 0 & 0 & \cdots \\ 1 & 0 & 0 & 0 & \cdots \\ 0 & 1 & 0 & 0 & \cdots \\ 0 & 0 & 1 & 0 & \cdots \\ \vdots & \vdots & \vdots & \ddots & \ddots \end{pmatrix}, \quad \widehat{S}_w^\dagger = \begin{pmatrix} 0 & 1 & 0 & 0 & \cdots \\ 0 & 0 & 1 & 0 & \cdots \\ 0 & 0 & 0 & 1 & \cdots \\ 0 & 0 & 0 & 0 & \ddots \\ \vdots & \vdots & \vdots & \vdots & \ddots \end{pmatrix}. \tag{S7}$$

We now solve the Schrödinger equation for Eq. (S6) by substituting the ansatz $|\psi_{\mathbf{k}}\rangle = \sum_{i=0}^{\infty} \lambda^i |i\rangle \otimes |\xi_{\mathbf{k}}\rangle$ with $|\lambda| < 1$ for boundary eigenstates. Inside the bulk with $|i \geq 1\rangle$, the Schrödinger equation gives

$$\mathcal{H}_{bulk}(\lambda)|\xi\rangle = \mathcal{E}|\xi\rangle, \tag{S8}$$

where

$$\mathcal{H}_{bulk}(\lambda) = \sum_{i=1}^{3} \sin k_i \Gamma_i + \frac{1}{2i}(\lambda^{-1} - \lambda)\Gamma_4 + \left( m - \sum_{i=1}^{3} \cos k_i - \frac{1}{2}(\lambda^{-1} + \lambda) \right)\Gamma_0, \tag{S9}$$

while at $|i = 0\rangle$

$$\mathcal{H}_{edge}(\lambda)|\xi\rangle = \mathcal{E}|\xi\rangle, \tag{S10}$$

with

$$\mathcal{H}_{edge}(\lambda) = \sum_{i=1}^{3} \sin k_i \Gamma_i - \frac{1}{2i}\lambda\Gamma_4 + \left( m - \sum_{i=1}^{3} \cos k_i - \frac{1}{2}\lambda \right)\Gamma_0. \tag{S11}$$

The difference of Eqs. (S8) and (S10) gives

$$i\Gamma_4\Gamma_0|\xi\rangle = |\xi\rangle. \tag{S12}$$

This equation implies the boundary states occur merely in the 4D subspace with positive eigenvalue 1 of $i\Gamma_4\Gamma_0$, which corresponds to the projector

$$\Pi = \frac{1}{2}(1 + i\Gamma_4\Gamma_0). \tag{S13}$$

Applying the projector to Eq. (S10), we have

$$\left( \sum_{i=1}^{3} \sin k_i \Gamma_i \right)|\xi\rangle = \mathcal{E}|\xi\rangle, \tag{S14}$$

The difference of Eqs. (S14) and (S10), together with Eq. (S12), gives

$$\lambda = m - \sum_i \cos k_i.$$ (S15)

The effective Hamiltonian for boundary states is just

$$\mathcal{H}_{eff}(\mathbf{k}) = \Pi \mathcal{H}(\mathbf{k}) \Pi = \left( \sum_{i=1}^{3} \sin k_i \Gamma_i \right) \Pi$$ (S16)

for regions in $\mathbf{k}$ space satisfying Eq. (S15).

More explicitly, in what follows we give the effective boundary Hamiltonian in the 4D subspace of $i\Gamma_4\Gamma_0$ with positive eigenvalues 1. After performing a unitary transformation $\mathcal{U}$ to diagonalize $i\Gamma_4\Gamma_0$ as $\sigma_3 \otimes \mathbf{1}_4$, we have

$$\mathcal{U} \Pi \mathcal{U}^{-1} = \begin{pmatrix} \mathbf{1}_4 & 0 \\ 0 & 0 \end{pmatrix},$$ (S17)

where $\mathbf{1}_4$ denotes the $4 \times 4$ identity matrix and

$$\mathcal{U} = e^{-i\frac{\pi}{4}\Gamma_4}.$$ (S18)

Employing the same procedure for effective Hamiltonian and taking the first non-zero 4D subspace, we have

$$\mathcal{H}_{bd} = \mathrm{Tr}\, \mathcal{U} \mathcal{H}_{eff} \mathcal{U}^{-1} = \sum_i \sin k_i \gamma^i,$$ (S19)

where "Tr" denotes the first non-zero 4D subspace. The gamma matrices are defined in the following formula

$$\mathcal{U} \Gamma_i \mathcal{U}^{-1} = \sigma_0 \otimes \gamma_i,$$ (S20)

where

$$\gamma_1 = -G_{13}, \ \gamma_2 = G_{30}, \ \gamma_3 = -G_{11},$$ (S21)

with $G_{ij} = \sigma_i \otimes \sigma_j$ as defined in the main text.

Note that the projectors for different boundaries perpendicular to $\alpha$-direction are given by

$$\Pi_{\pm}^{\alpha} = \frac{1}{2} \left( 1 \pm i\Gamma_\alpha \Gamma_0 \right),$$ (S22)

where $\alpha = x, y, z, w$, and $\pm$ denotes the boundaries at $L = 0$ and $L = L_\alpha$, respectively. The corresponding effective boundary Hamiltonian in the 4D subspace associates to $i\Gamma_\alpha\Gamma_0$ with eigenvalues $\pm 1$.

For convenience, we present all the 3D boundary Hamiltonian (that fall into Class DIII) when considering a boundary normal to the $i$-direction, which are,

$$\begin{aligned}
\mathcal{H}_{eff}^{x,\pm} &= \pm (\sin k_y G_{23} + \sin k_z G_{02} + \sin k_w G_{13}), \quad (\mathcal{T} = G_{12}\mathcal{K}, \mathcal{C} = G_{13}\mathcal{K}, \mathcal{S} = G_{01}) \\
\mathcal{H}_{eff}^{y,\pm} &= \mp (\sin k_x G_{23} - \sin k_z G_{21} + \sin k_w G_{30}), \quad (\mathcal{T} = G_{12}\mathcal{K}, \mathcal{C} = G_{30}\mathcal{K}, \mathcal{S} = G_{22}) \\
\mathcal{H}_{eff}^{z,\pm} &= \mp (\sin k_x G_{02} + \sin k_y G_{21} + \sin k_w G_{11}), \quad (\mathcal{T} = G_{12}\mathcal{K}, \mathcal{C} = G_{11}\mathcal{K}, \mathcal{S} = G_{03}) \\
\mathcal{H}_{eff}^{w,\pm} &= \mp (\sin k_x G_{13} - \sin k_y G_{30} + \sin k_z G_{11}), \quad (\mathcal{T} = G_{12}\mathcal{K}, \mathcal{C} = \mathcal{K}, \mathcal{S} = G_{12})
\end{aligned}$$ (S23)

where $G_{ij} = \sigma_i \otimes \sigma_j$. All these Hamiltonian commutes with $G_{32}$. These boundary Hamiltonian represents the 3D Dirac points carry a $\mathbb{Z}_2$ monopole.

## A. Effective Boundary Hamiltonian for Model $\mathcal{H}_1$

After introducing $b_\mu$ term, the corresponding boundary Hamiltonian of the total Hamiltonian $\mathcal{H}_1$ takes the form,

$$\begin{aligned}
\mathcal{H}_1^{x,\pm} &= \mathcal{H}_{eff}^{x,\pm} \pm b_2 G_{11} \pm b_3 G_{30}, \\
\mathcal{H}_1^{y,\pm} &= \mathcal{H}_{eff}^{y,\pm} \mp b_1 G_{11} \pm b_3 G_{13}, \\
\mathcal{H}_1^{z,\pm} &= \mathcal{H}_{eff}^{z,\pm} \pm b_1 G_{30} \pm b_2 G_{13} \pm b_0 G_{32}, \\
\mathcal{H}_1^{w,\pm} &= \mathcal{H}_{eff}^{w,\pm} \pm b_1 G_{21} \pm b_2 G_{02} \pm b_3 G_{23} \pm b_0 G_{32},
\end{aligned}$$ (S24)

where "+"("-") denotes the boundary at $L = 0 (L_\alpha)$.

## B. Effective Boundary Hamiltonian for Model $\mathcal{H}_2$

In the following, we calculate the effective boundary Hamiltonian for $\mathcal{H}_2$ in Eq. (S48) in Sec. S-V A. When $\delta_z$ is small before closing the band gap, the boundary effective Hamiltonian is the same as in Eq. (S23). After the topological phase transition, i.e., the bulk gap closes and reopens, the topology of each block Hamiltonian is characterized by different parameter $m_\pm$. For convenience, we discuss the effective boundary Hamiltonian of block $\mathcal{H}_2^+$ and $\mathcal{H}_2^-$, respectively.

The projector upon applying open boundary condition along $w$-direction at $L = 0$ for each sub system is given by

$$\Pi_\pm = \frac{1}{2}\left(1 \mp iG_{20}G_{32}\right). \tag{S25}$$

Similarly, we obtain the effective boundary Hamiltonian after performing a unitary transformation and take the first non-zero subspace,

$$\begin{aligned}\mathcal{H}_{bd}^\pm &= \mathrm{Tr}\,\mathcal{U}_\pm\mathcal{H}_2^\pm\mathcal{U}_\pm^{-1} \\ &= -(\sin k_x\sigma_3 \mp \sin k_y\sigma_2 + \sin k_z\sigma_1),\end{aligned} \tag{S26}$$

with $\mathcal{U}_\pm = \exp\left(\pm i\frac{\pi}{4}G_{22}\right)$ under the condition

$$\lambda_\pm = m_\pm - \sum_{i=x,y,z}\cos k_i. \tag{S27}$$

We emphasize again that the above effective Hamiltonian only survives and supports the boundary states when $|\lambda_\pm| < 1$.

For the original case with $m = 3$ and $\delta_z = 0$, the sub system in Eq. (S26) represents a Weyl point at $(0,0,0)$ which can be expanded as

$$\mathcal{H}_W^\pm = -(k_x\sigma_3 \mp k_y\sigma_2 + k_z\sigma_1) \tag{S28}$$

with opposite chirality $\eta = \mp 1$, respectively. At this time, $\lambda_\pm = 0$. Increasing $\delta_z$ until 1, the band gap of $\mathcal{H}_+$ [$\mathcal{H}_-$] closes at $(0,0,0,0)$ [$(0,0,\pi,0)$, $(0,\pi,0,0)$, $(\pi,0,0,0)$] where the corresponding $\lambda_+ = 1$ [$\lambda_- = 1$]. In the later process by enlarging $\delta_z$, $|\lambda_+|$ is always greater than 1 and thus has no contributions to the boundary states while $|\lambda_-|$ is less than 1 until $\delta_z > 7$. Repeating the above process with different $m$ and $\delta_z$, we obtain the full phase diagram and the information of the boundary Weyl points which is consistent with the bulk topological number as we discussed below in the Sec. S-V A.

## S-III. (3+1)D BOUNDARY RESPONSES

It is known that a Weyl fermion coupled to a gauge field leading to axial anomaly, which could be described by the anomaly equation[6, 7],

$$\partial_\mu j^{\mu(\pm)} = \pm\frac{1}{32\pi^2}\epsilon^{\mu\nu\rho\sigma}F_{\mu\nu}^{(\pm)}F_{\rho\sigma}^{(\pm)}, \tag{S29}$$

where $\pm$ denotes right- and left-handed Weyl fermion, respectively. Replacing the new gauge field strength, i.e.,

$$F_{\mu\nu} = \frac{1}{2}\left[F_{\mu\nu}^{(+)} + F_{\mu\nu}^{(-)}\right],\ \tilde{F}_{\mu\nu} = \frac{1}{2}\left[F_{\mu\nu}^{(+)} - F_{\mu\nu}^{(-)}\right], \tag{S30}$$

into Eq. (S29), and then defining the charge and axial currents as $j^\mu = j^{\mu(+)} + j^{\mu(-)}$, and $\tilde{j}^\mu = j^{\mu(+)} - j^{\mu(-)}$ respectively. Thus we have

$$\begin{aligned}\partial_\mu j^\mu &= \frac{1}{8\pi^2}\epsilon^{\mu\nu\rho\sigma}\tilde{F}_{\mu\nu}F_{\rho\sigma}, \\ \partial_\mu\tilde{j}^\mu &= \frac{1}{16\pi^2}\epsilon^{\mu\nu\lambda\rho\sigma}\left(F_{\nu\lambda}F_{\rho\sigma} + \tilde{F}_{\nu\lambda}\tilde{F}_{\rho\sigma}\right),\end{aligned} \tag{S31}$$

Therefore, the anomaly equation of the 3D boundary Driac fermions of a 4D topological insulator can be expressed as

$$\sum_i \partial_\mu \hat{j}_i^\mu = \frac{C_{2s}}{8\pi^2} \epsilon^{\mu\nu\rho\sigma} \tilde{F}_{\mu\nu} F_{\rho\sigma},$$

$$\sum_i \partial_\mu \tilde{j}_i^\mu = \frac{C_{2s}}{16\pi^2} \epsilon^{\mu\nu\rho\sigma} \left( F_{\mu\nu} F_{\rho\sigma} + \tilde{F}_{\mu\nu} \tilde{F}_{\rho\sigma} \right),$$

(S32)

where $i$ denotes the number of Dirac fermions where each of them consists of a pair of Weyl fermions with opposite chirality on the 3D boundary. $C_{2s} = \Delta C_2/2$ denotes the number of Dirac fermions. Thus, the response current density from the boundary is consistent with the result discussed from the bulk when we take the same gauge as in the main text.

On the other hand, when we introduce $b_\mu$-field instead of spin gauge field $\tilde{A}_\mu$. The response currents can be derived following the same procedure, which takes the form,

$$\sum_i \partial_\mu j_i^\mu = \frac{C_{2v}}{8\pi^2} \epsilon^{\mu\nu\rho\sigma} f_{\mu\nu} F_{\rho\sigma},$$

$$\sum_i \partial_\mu j_{5,i}^\mu = \frac{C_{2v}}{16\pi^2} \epsilon^{\mu\nu\rho\sigma} \left( F_{\mu\nu} F_{\rho\sigma} + f_{\mu\nu} f_{\rho\sigma} \right),$$

(S33)

where $i$ denotes how many pairs of Weyl points are on the 3D boundary. The first formula denotes the regular topological charge current in a 3D Weyl semimetal while the second formula denotes the axial current when $b_\mu$ is a strain-induced gauge field[5].

In addition, we still have the bulk-edge correspondence from viewpoint of the transport. However, when $b_\mu$ is a constant field, we only obtain the non-zero valley current $J_5^\mu$ in the 4D bulk. In contrast, the topological charge current are non-zero on the 3D boundary, which are,

$$j^\mu = \sum_i j_i^\mu = \frac{C_{2v}}{8\pi^2} \epsilon^{\mu\nu\rho\sigma} b_\nu F_{\rho\sigma},$$

(S34)

where $j^\mu = (\rho_0, \mathbf{j})$ with $\rho_0$ denotes particle density and $\mathbf{j}$ is the charge current density. When only $b_0$ survives, two Weyl points separated in energy form a nodal sphere structure and the corresponding response current represents the chiral magnetic effect, i.e., $\mathbf{j} \propto b_0 \mathbf{B}$; while only $\mathbf{b}$ is non-zero, we have the quantum anomalous hall effect $\mathbf{j} \propto \mathbf{b} \times \mathbf{E}$ originates from the 2D sub systems between these two Weyl nodes separated in momentum.

## S-IV. FIELD-THEORETICAL DESCRIPTION OF A (2+1)D $\mathbb{Z}_2$ TOPOLOGICAL INSULATOR

We start by considering a well-known 2D time-reversal-invariant insulator in the absence of spin-orbit-coupling that falls into class DIII. Its Hamiltonian takes the form

$$\mathcal{H}_{2D} = \begin{pmatrix} h_+(k) & 0 \\ 0 & h_-(k) \end{pmatrix},$$

(S35)

where

$$h_\pm(k) = \pm \sin k_x \sigma_1 + \sin k_y \sigma_2 + (m - \cos k_x - \cos k_y)\sigma_3.$$

(S36)

Each block subsystem $h_\pm$ represents a Chern insulator that hosts a first Chern number (CN),

$$C_1^\pm = \begin{cases} \mp \text{sgn}(m), & 0 < |m| < 2, \\ 0, & |m| > 2. \end{cases}$$

(S37)

In the meantime, we can define $\nu_1 = C_{1s} \bmod 2$ as the $\mathbb{Z}_2$ invariant for this system where the first spin CN $C_{1s} = (C_1^+ - C_1^-)/2$. In the following part, we always mention on the case while the system commutes with $s_z = G_{30}$, i.e., $s_z$ is conserved. We consider adding a TRS-broken perturbation

$$\Delta_s = b_0 G_{30} + b_1 G_{01} + b_2 G_{32}$$

(S38)

into $\mathcal{H}_{2D}$. This term will lift the band degeneracy of $\mathcal{H}_{2D}$ and shift two band valleys in energy and momentum with difference $(\delta E, \delta k_x, \delta k_y) = (2b_0, 2b_1, 2b_2)$. Thus, the $\mathbb{Z}_2$ nature of the whole system remains unchanged for a small $\Delta_s$. The corresponding edge Hamiltonian with an open boundary normal to $x$ direction is now given by

$$\mathcal{H}_{edge} = \begin{pmatrix} -(\sin k_y - b_2) - b_0 & 0 \\ 0 & \sin k_y + b_2 + b_0 \end{pmatrix}. \tag{S39}$$

We can see that $b_\mu$ also shifts two opposite-handed chiral fermions in energy-momentum space.

Similar to the idea in main text, each 2D block system coupled to a gauge field $A_\mu^{(\pm)}$ is described by the (2+1)D Chern-Simons theory,

$$S_{cs}^{(\pm)} = \frac{C_1^\pm}{4\pi} \int d^3x \epsilon^{\mu\nu\lambda} A_\mu^{(\pm)} \partial_\nu A_\lambda^{(\pm)}. \tag{S40}$$

Thus, for the case $b_\mu = 0$, we introduce a spin gauge field $\tilde{A}_\mu = \frac{1}{2} \left[ A^{(+)} - A^{(-)} \right]$, then obtain the mixed Chern-Simons action

$$S_{s,eff} = \frac{C_{1s}}{\pi} \int d^3x \epsilon^{\mu\nu\lambda} \tilde{A}_\mu \partial_\nu A_\lambda. \tag{S41}$$

The corresponding responses

$$J^\mu = \frac{C_{1s}}{\pi} \epsilon^{\mu\nu\lambda} \partial_\nu \tilde{A}_\lambda, \ \tilde{J}^\mu = \frac{C_{1s}}{\pi} \epsilon^{\mu\nu\lambda} \partial_\nu A_\lambda. \tag{S42}$$

We here pick the gauge

$$A_\mu = (0, 0, -tE_y), \ \tilde{A}_\mu = 0, \tag{S43}$$

then we have

$$J^x = 0, \ \tilde{J}^x = \frac{C_{1s}}{\pi} E_y. \tag{S44}$$

Thus $\tilde{J}^x$ represents the spin hall effect with a quantized spin conductance $C_{1s}/\pi$.

Likewise, when we consider a non-zero $b_\mu$-field, the effective action is now given by

$$S_{v,eff} = \frac{C_{1v}}{\pi} \int d^3x \epsilon^{\mu\nu\lambda} b_\mu \partial_\nu A_\lambda, \tag{S45}$$

where the first valley CN $C_{1v} = C_{1s}$ with the corresponding charge and valley currents

$$J^\mu = \frac{C_{1v}}{\pi} \epsilon^{\mu\nu\lambda} \partial_\nu b_\lambda, \ J_5^\mu = \frac{C_{1v}}{\pi} \epsilon^{\mu\nu\lambda} \partial_\nu A_\lambda. \tag{S46}$$

Similarly, picking $b_\mu = (b_0, 0, -b_y(t))$, then we have

$$J^x = \frac{C_{1v}}{\pi} E_y^5, \ J_5^x = \frac{C_{1v}}{\pi} E_y, \tag{S47}$$

where the pseudo-electric field $E_y^5 = \partial_t b_y$. Note that $J^x$ denotes the valley-induced effect(say, "*pseudo-Hall effect*") with a quantized hall conductivity $C_{1v}/\pi$ while $J_5^x$ denotes the valley hall effect with a quantized valley hall conductivity $C_{1v}/\pi$.

In addition, one can easily obtain the corresponding edge response arising from two chiral fermions with opposite chirality which is consistent with the response from the bulk. In summary, we can say that the above (2+1)D effects can be covered through the dimensional reduction from the (4+1) D case as we did in the main text.

## S-V. TOPOLOGICAL PHASE TRANSITIONS

### A. Insulator-Insulator transition

In this section, we move to discuss the topological phase transition after introducing an extra term $\Delta_z = \delta_z G_{032}$ which only preserves TRS but breaks all the rest of symmetries. The model will still host a $\mathbb{Z}_2$ Dirac cone on the

3D boundary until the bulk gap closes and reopens at a critical value $\delta_c$ since we tune this parameter from 0. After that, the system falls into class AII and thus is characterized by a $\mathbb{Z}$ number which is nothing but the second Chern number. In the following, we will interpret this picture through analysing the bulk and boundary physics for the concrete model.

Since such an extra term $\delta_z G_{032}$ is still diagonalized after the rotation, the total Hamiltonian in the block representation now is

$$\mathcal{H}_2 = \mathcal{H}_{BD} + \delta_z G_{032} = \begin{pmatrix} \mathcal{H}_2^+ & 0 \\ 0 & \mathcal{H}_2^- \end{pmatrix} \tag{S48}$$

with each block Hamiltonian

$$\mathcal{H}_2^\pm = \mathcal{H}_\pm + \delta_z G_{32}. \tag{S49}$$

The energy spectrum takes the form,

$$
\begin{aligned}
E_+ &= \pm\sqrt{d_x^2 + d_y^2 + d_z^2 + d_w^2 + (d_m + \delta_z)^2}, \\
E_- &= \pm\sqrt{d_x^2 + d_y^2 + d_z^2 + d_w^2 + (d_m - \delta_z)^2},
\end{aligned}
\tag{S50}
$$

where $E_+$ and $E_-$ denote the energy spectrum of $\mathcal{H}_2^+$ and $\mathcal{H}_2^-$, respectively. As we can find that each block model is just to be modified by replacing $m$ to $m_\pm = m \pm \delta_z$. This means that we can directly obtain the second Chern number of each block by just replacing $m$ with $m_\pm$ in $\mathcal{H}_\pm$. Meanwhile, the phase transition points should be modified as

$$
\begin{aligned}
m + \delta_z &= 0, \pm2, \pm4 \\
m - \delta_z &= 0, \pm2, \pm4,
\end{aligned}
\tag{S51}
$$

which leading to ten phase transition boundaries

$$
\begin{aligned}
\delta_{0,+} &= -m, \delta_{1,\pm} = -m \pm 2, \delta_{2,\pm} = -m \pm 4, \\
\delta_{0,-} &= +m, \delta_{3,\pm} = +m \mp 2, \delta_{4,\pm} = +m \mp 4.
\end{aligned}
\tag{S52}
$$

The phase diagram is now confirmed, as shown in Fig. S1.

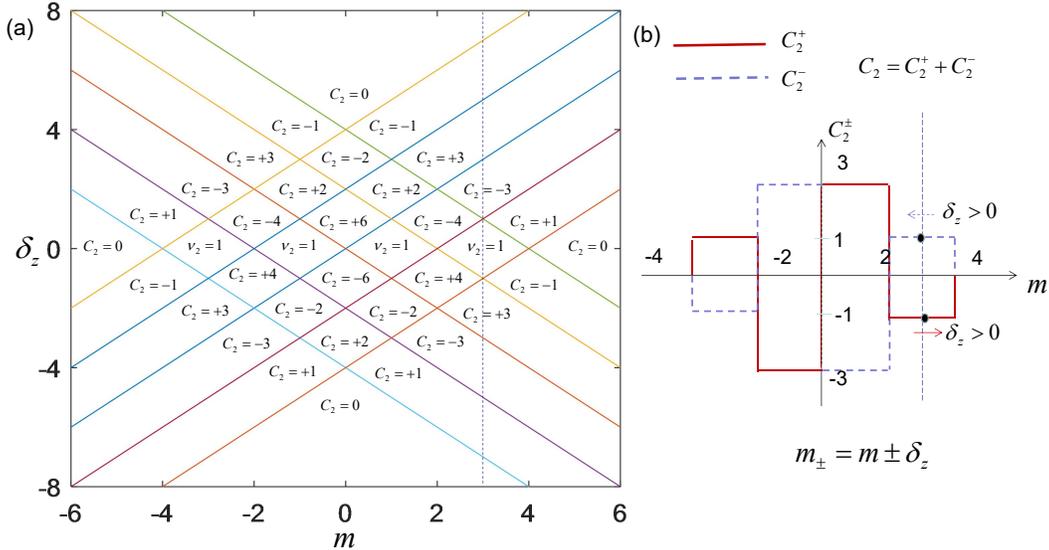

FIG. S1. (a)Schematic of the phase diagram of $\mathcal{H}_2$ with parameters $(m, \delta_z)$. The straight color lines denote the phase transition boundaries where the bulk gap closes at $E = 0$. Note that $C_2 = C_2^+ + C_2^-$, and the purple dashed line is at the value $m = 3$. (b) The schematic of the phase diagram of $\mathcal{H}_\pm$ with $C_2^\pm$ by varying $m$ and $\delta_z$.

Because the phase transition points for each subsystem are no longer synchronizing when $\delta_z$ is nonzero, it leads to a very rich phase diagram. To explain why we have such a phase diagram, we discuss the phase transition for

a concrete case at a value $m = 3$ by varying $\delta_z$. When $\delta_z = 0$, this is a non-trivial $\mathbb{Z}_2$ topological insulator with $C_\pm = \mp 1$, i.e., $\nu_2 = 1$, as we discussed before. The system now hosts a $\mathbb{Z}_2$ Dirac cone consists of two Weyl cones with opposite chirality on the 3D boundary and is in class AII. When we increase $\delta_z$ from 0, the 3D boundary mode is still there until the bulk gap for $\mathcal{H}_2^\pm$ closes at $\delta_z = 1$. Even the system breaks all the symmetries except TRS, $\nu_2$ is still well-defined. The reason is that each block Hamiltonian $\mathcal{H}_2^\pm$ belongs to class A is quite robust and thus still hosts the unchanged second Chern number $C_2^\pm = \mp 1$ after adding a perturbation $\pm \delta_z G_{30}$. Increasing $\delta_z$, the bulk gap reopens, however, the parameter $m_+ > 4$ of $\mathcal{H}_2^+$ falls into the trivial range hereafter and thus this block has no contributions to the 3D boundary states. In contrast, $\mathcal{H}_2^-$ hosts a second Chern number $C_2^- = -3$ and supports three left-handed Weyl fermions at $(0, 0, \pi)$, $(0, \pi, 0)$, and $(\pi, 0, 0)$ respectively within the 3D boundary BZ. Here we consider the open boundary condition along $w$-direction. When reaching $\delta_z = 3$ and later, band gap closes and reopens, $\mathcal{H}_2^-$ has $C_2^- = +3$ and supports three right-handed Weyl fermions at $(0, \pi, \pi)$, $(\pi, \pi, 0)$, and $(\pi, 0, \pi)$, respectively. If we further increase $\delta_z$ to be greater than 5, one obtain $C_2^- = -1$ and the corresponding Weyl fermion at $(\pi, \pi, \pi)$. After the last phase transition at $\delta_z = 7$, the whole system becomes totally trivial with $C_2 = 0$. Following this analysis process with varying $m$ and $\delta_z$, we give a complete phase diagram, as depicted in Fig. S1. Alternatively, one can analyze this process from its 3D boundary physics as presented in Sec. S-II B above.

It is worth noticing that the system remains to host the $\mathbb{Z}_2$ nature even we break all the symmetries except $\mathcal{T}$. In principle, this model has already fallen into class AII and should be characterized by a $\mathbb{Z}$ number. However, since the robustness of the topology for each block model that is in class A, the bulk second CN and boundary states of each block system will not change until the bulk gap closes and reopens. This picture is similar to adding a TRS-broken term into the 2D TRI topological insulators where the system still preserves its $\mathbb{Z}_2$ nature until the gap closes and reopens[8]. That is to say, the topological phase transition from $\mathbb{Z}_2$ to $\mathbb{Z}$ insulators can not happen without closing the band gap. After this phase transition, the system really falls into class AII and becomes a 4D quantum Hall insulator. It hosts a total second Chern number $C_2 = C_2^+ + C_2^-$ with $C_2^+$ and $C_2^-$ determined by different parameter $m_+$ and $m_-$, respectively. This is why we obtain several extra values $C_2 = \pm 2, \pm 4, \pm 6$ in Fig. S1.

## B. Insulator-Metal-Insulator transition

### 1. Phase transition

We now consider adding a term $\Delta_a = c G_{302} + b_0 G_{332}$ to break all the symmetries so that the model falls into class A and is now given by

$$\mathcal{H}_3 = \mathcal{H}_0 + \Delta_a,\tag{S53}$$

with the spectrum

$$E = \pm b_0 \pm \sqrt{d_y^2 + d_w^2 + \left(\sqrt{d_x^2 + d_z^2 + d_m^2} \pm c\right)^2}.\tag{S54}$$

For simplicity, we first consider the simple case with $b_0 = 0$. When we increase (decrease) $c$ from 0, the system still in the $\mathbb{Z}_2$ phase until the band gap closes and reopens. After that, we obtain double-degenerate nodal line structure for the two middle degenerate bands and then finally becomes a trivial insulator when $c$ is large (small) enough. During this process, the double nodal line structure stems from the middle two band crossing is determined by

$$d_y = d_w = 0 \Rightarrow k_y, k_w = 0, \pi,$$
$$d_x^2 + d_z^2 + d_m^2 = c^2 \Rightarrow \sin k_x^2 + \sin k_z^2 + (m - \sum_i \cos k_i)^2 = c^2.\tag{S55}$$

For instance, we start by considering the case when $m = 3$ and $c = 0$. Increasing $c$, the system is still a $\mathbb{Z}_2$ TI until $c = 1$ when the bulk gap closes. In the region $1 < c < 3$, there are three double-nodal-line formed by the middle two bands with the centers $k_0$ at $k_x$-$k_z$ plane when $(k_y, k_w) = (k_y^c, k_w^c)$,

$$(k_y, k_w) = (0, 0), k_0 = (\pi, \pi),$$
$$(k_y, k_w) = (0, \pi), k_0 = (0, 0),$$
$$(k_y, k_w) = (\pi, 0), k_0 = (0, 0).\tag{S56}$$

To further enlarge $c$ where $3 < c < 5$, the three double-nodal-lines are now located at

$$
\begin{aligned}
(k_y, k_w) &= (0, \pi), k_0 = (\pi, \pi), \\
(k_y, k_w) &= (\pi, 0), k_0 = (\pi, \pi), \\
(k_y, k_w) &= (\pi, \pi), k_0 = (0, 0).
\end{aligned}
\tag{S57}
$$

In the region $5 < c < 7$, there is one double-nodal-line surviving at $k_x$-$k_z$ plane with $k_0 = (\pi, \pi)$ when $k_y = k_w = \pi$. After that, the band gap opens and becomes a trivial insulator when $c > 7$. After identifying the information of band structure with varying $m$ and $c$, we obtain a phase diagram as depicted in Fig. S2(a). The phase boundaries are the same as in $\mathcal{H}_2$ but the all the non-trivial Chern phases are replaced by gapless double-nodal-line semimetallic phases. Another interesting thing is that each double-nodal-line will split in energy and become a nodal-hyper-torus ($\mathbb{T}^3$) once $b_0$ takes a small value. This 3D torus can be treated as an inflated double-nodal-line.

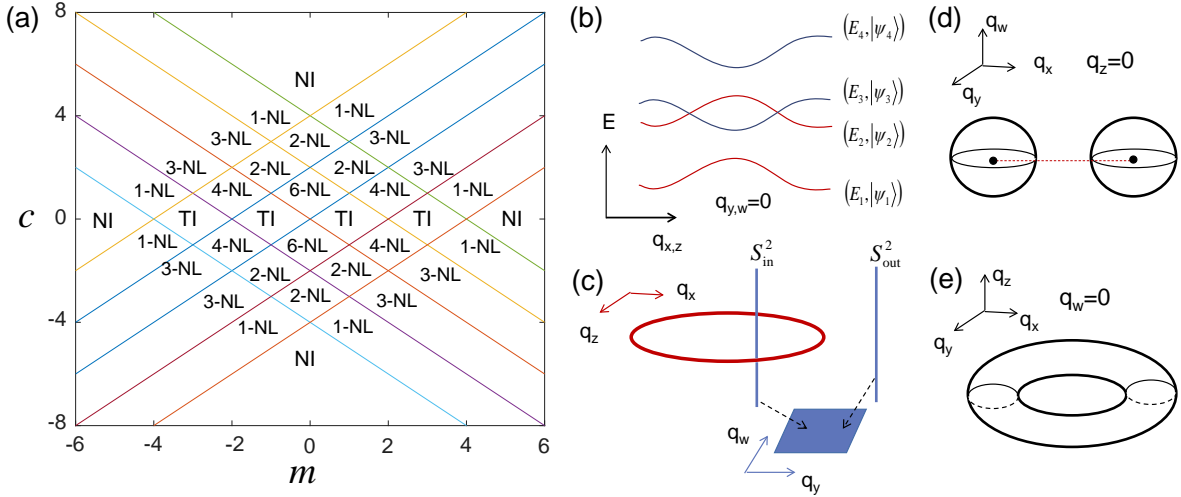

FIG. S2. (a)Schematic of the phase diagram of $\mathcal{H}_3$ with parameters $(m, c)$. The straight color lines denote the phase transition boundaries where the bulk gap closes at $E = 0$. Note that "TI" denotes $\mathbb{Z}_2$ TIs while "NI" denotes normal insulators like in Fig. S1. "$i$-NL" denotes the system hosts $i$ double nodal lines forming by the middle two degenerate bands. (b) Schematic of the energy spectrum of $\mathcal{H}^+_{0,eff}$ when $q_{y,w} = 0$. (c) Schematic of the double-nodal-line structure. Cross-section of a nodal-hyper-torus $\mathbb{T}^3$:(d) A pairs of nodal spheres when $q_z = 0$; (e) A single torus $\mathbb{T}^2$ when $q_w = 0$.

### 2. Topology of a double-nodal-line/nodal-hyper-torus

Without loss of generality, we consider the case when $5 < c < 7$ and $m = 3$ with $b_0 = 0$. The system supports one double-nodal-line expanded around $k_c = (\pi, \pi, \pi, \pi)$ satisfying the equation: $q_x^2 + q_z^2 = \rho_0^2$ with the radius $\rho_0 = \sqrt{c^2 - m^2}$ when $k_{y,w} = \pi$. For convenience, we consider rotating $\mathcal{H}_3$ into the block diagonal representation, i.e.,

$$
U \mathcal{H}_3 U^{-1} = \begin{pmatrix} \mathcal{H}_+ - cG_{30} & 0 \\ 0 & \mathcal{H}_- + cG_{30} \end{pmatrix}.
\tag{S58}
$$

There is a nodal-line structure for each block with the low-energy Hamiltonian around $k_1^c$ given by

$$
H^\pm_{eff} = H^\pm_{0,eff} + V_\pm,
\tag{S59}
$$

where

$$
\begin{aligned}
H^\pm_{0,eff} &= q_x G_{33} + q_z G_{31} \pm \tilde{m} G_{32} \mp cG_{30} \\
V_\pm &= q_y G_{10} - q_w G_{20},
\end{aligned}
\tag{S60}
$$

with $\tilde{m} = m + q_\parallel^2/2$ and $q_\parallel = \sqrt{q_y^2 + q_w^2}$. For $\mathcal{H}_{eff}^\pm$, there is a nodal-line structure formed by the middle two bands for each subsystem, thus we use the degenerate perturbation theory to calculate the two-level effective Hamiltonian by considering $V_\pm$ as a perturbation.

Take $\mathcal{H}_{0,eff}^+$ as an example, its spectrum $E = \pm c \pm \sqrt{q_x^2 + q_z^2 + \tilde{m}^2}$ as shown in Fig. S2(b) when $q_y = q_w = 0$. We label these bands as $E_i$ with $E_i < E_j$ for $i = 1, 2, 3, 4$. Now we apply the degenerate perturbation theory for the middle two bands, i.e., $(E_2, E_3)$, with the eigenstates $(|\psi_2\rangle, |\psi_3\rangle)$. After straightforward calculation into the first-order, i.e., $(\mathcal{H}_{NL}^+)_{ij} = \langle \psi_i | (\mathcal{H}_{0,eff}^+ + V_+) | \psi_j \rangle$, we obtain

$$\mathcal{H}_{NL}^+ = -q_w \sigma_1 + q_y \sigma_2 + (\sqrt{q_\perp^2 + \tilde{m}^2} - c)\sigma_3, \tag{S61}$$

where $q_\perp = \sqrt{q_x^2 + q_z^2}$. One can see that, when $q_\perp < \rho_0$ inside the nodal line, this model represents the continuum model of a nontrivial 2D Chern insulator in $\mathbf{q}_\parallel = (q_y, q_w)$ space. Such as, for $q_\perp = 0$, we have,

$$\mathcal{H}_{CI} = -q_w \sigma_1 + q_y \sigma_2 + (m - c + \frac{q_\parallel^2}{2})\sigma_3, \tag{S62}$$

with the first CN is given by

$$C_1^+ = \frac{1}{2\pi} \int_{\mathbb{R}^2} dq_y dq_w \mathcal{F}_{yw}^+ = \frac{1 + \text{sgn}(c - m)}{2} = 1, \tag{S63}$$

while the subsystem $\mathcal{H}_{q_\perp > \rho_0}(q_y, q_w)$ is a trivial insulator with $C_1^+ = 0$. Thus, the subsystem $\mathcal{H}_{q_\perp}(q_y, q_w)$ inside (outside) the nodal line $q_\perp < \rho_0$ $(q_\perp > \rho_0)$ is a nontrivial (trivial) Chern insulator which supports one (zero) chiral edge mode at each side when considering open boundary condition along $w$-direction.

Performing the similar calculation for $\mathcal{H}_{eff}^-$, we obtain the effective Hamiltonian given by

$$\mathcal{H}_{NL}^- = q_w \sigma_1 - q_y \sigma_2 + (\sqrt{q_\perp^2 + \tilde{m}^2} - c)\sigma_3, \tag{S64}$$

Note that this model shares the same topological features with $\mathcal{H}_{NL}^+$. Therefore, the total effective model of this double-nodal-line semimetal is given by

$$\mathcal{H}_{DNL} = \mathcal{H}_{NL}^+ \oplus \mathcal{H}_{NL}^- = -q_w G_{31} + q_y G_{32} + (\sqrt{q_\perp^2 + \tilde{m}^2} - c)G_{03}. \tag{S65}$$

This model also falls into the class A characterized by a $d_c = 2$D topological invariant with the codimensions $d_c = d - d_{FS} - 1$[9], the system dimensions $d = 4$ and the dimensions of gapless Fermi surface $d_{FS} = 1$. So we define a $\mathbb{Z}$ number as

$$\nu_z = C_1^{in} - C_1^{out} = 2, \tag{S66}$$

where

$$C_1 = \frac{1}{2\pi} \int_{\mathbb{R}^2}^{q_\perp = q_0} dq_y dq_w \text{tr} \mathcal{F}_{yw}, \tag{S67}$$

with $\text{tr} \mathcal{F}_{yw} = \mathcal{F}_{yw}^+ + \mathcal{F}_{yw}^-$ for the lower two-fold degenerate bands.

Once $b_0$ is non-zero, the $4 \times 4$ effective Hamiltonian is given by

$$\mathcal{H}_{NHT} = \mathcal{H}_{DNL} + b_0 G_{30}, \tag{S68}$$

with the spectrum

$$E = \pm b_0 \pm \sqrt{q_y^2 + q_w^2 + (\sqrt{q_\perp^2 + m^2} - c)^2}. \tag{S69}$$

Above we abandoned the higher-order terms $O(q_\parallel^n)$. It has a zero-energy band degeneracy at

$$q_y^2 + q_w^2 + (\sqrt{q_\perp^2 + m^2} - c)^2 = b_0^2. \tag{S70}$$

forming a nodal-hyper-torus in momentum space for $b_0 < \rho_0$. The dimensions of this nodal-hyper-torus is 3D which is nothing but $\mathbb{T}^3 = \mathbb{S}^1 \times \mathbb{S}^1 \times \mathbb{S}^1$. Taking some sections of this object, we can set either $q_z = 0$, or $q_w = 0$, then the cross-section

$$q_y^2 + q_w^2 + (\sqrt{q_x^2 + m^2} - c)^2 = b_0^2, \tag{S71}$$

gives us the disjoint pair of spheres centered at $(+\rho_0, 0, 0)$ and $(-\rho_0, 0, 0)$; or

$$q_y^2 + (\sqrt{q_\perp^2 + m^2} - c)^2 = b_0^2, \tag{S72}$$

which is nothing but a single torus $\mathbb{T}^2$. Thus the codimensions of this nodal-hyper-torus is $d_c = 0$. Since this model (S68) preserves the $\gamma_5$-symmetry, i.e, $[\gamma_5, \mathcal{H}_{NHT}] = 0$ with $\gamma_5 = G_{30}$. Thus we define a 0D $\mathbb{Z}$ number,

$$\nu_0 = \langle \psi(q) | \gamma_5 | \psi(q) \rangle, \tag{S73}$$

where $|\psi(q)\rangle$ denotes the occupied band with the highest energy. The topological invariant of this nodal-hyper-torus can be defined as $\Delta\nu_0 = [\nu_0(q_{in}) - \nu_0(q_{out})]/2 = 1$, where the momentum $q_{in(out)}$ is located anywhere inside (outside) the nodal-hyper-torus surface[10].

## S-VI. 4D $\mathbb{Z}_2$ TI MODELS IN CLASS CII AND C

In this section, we present the 4D minimal models in class CII and C. We below only mention on the systems that exhibit the predicted responses as discussed in the main text. Thus all these models should exist a conserved quantity, i.e., these systems should commute with a specific matrix. For convenience, we consider constructing these models in the block diagonal representation below. One can simply introduce a spin-orbit-coupling-like symmetry-protected term for further investigations.

### A. A minimal model in Class CII

For a 4D TI that falls into class CII, Hamiltonian $\mathcal{H}_{CII}$ should preserve the time-reversal symmetry(TRS), partial-hole symmetry (PHS), and chiral symmetry (CS) with the operators

$$\begin{aligned}
\mathcal{T}\mathcal{H}(k)\mathcal{T}^{-1} &= \mathcal{H}(-k), \mathcal{T} = U_t\mathcal{K}, \ \mathcal{T}^2 = -1, \\
\mathcal{C}\mathcal{H}(k)\mathcal{C}^{-1} &= -\mathcal{H}(-k), \mathcal{C} = U_c\mathcal{K}, \ \mathcal{C}^2 = -1, \\
\mathcal{S}\mathcal{H}(k)\mathcal{S}^{-1} &= -\mathcal{H}(k), \mathcal{S} = U_tU_c^*, \ \mathcal{S}^2 = +1.
\end{aligned} \tag{S74}$$

Since $\mathcal{T}^2 = -1$, thus this is a spinful system, e.g., spin-1/2. For simplicity, we consider a minimal model of 4D $\mathbb{Z}_2$ TI consists of two copies of 4D time-reversal-invariant (TRI) quantum Hall insulators (QHIs) with opposite odd second Chern numbers (SCNs),i.e.,

$$\mathcal{H}_{CII}(k) = \begin{pmatrix} \mathcal{H}_{AII}^+(k) & 0 \\ 0 & \mathcal{H}_{AII}^-(k) \end{pmatrix}, \tag{S75}$$

where $\mathcal{H}_{AII}^+$ is the well-known 4D TRI insulator and takes the form,

$$\mathcal{H}_{AII}^+ = d_x G_{31} + d_y G_{32} + d_z G_{33} + d_w G_{20} + d_m G_{10}, \tag{S76}$$

with $G_{ij}$ and the Bloch vector are defined as in the main text. Here the Bloch vector components satisfy

$$d_i(k) = -d_i(-k), \forall i = x, y, z, w, \ d_m(k) = d_m(-k). \tag{S77}$$

The TR operator of this model is $\mathcal{T}_+ = iG_{02}\mathcal{K}$ with $\mathcal{T}_+^2 = -1$ and thus belongs to class AII. If we take $\mathcal{H}_{AII}^- = -\mathcal{H}_{AII}^+$, then this sub system hosts an opposite SCN to $\mathcal{H}_{AII}^+$ and also preserves TRS with $\mathcal{T}_- = \mathcal{T}_+$. Therefore, the total model $\mathcal{H}_{CII}$ preserves TRS, PHS, and CS. We can find one representation of the operators is

$$\begin{aligned}
\mathcal{T} &= iG_{002}\mathcal{K}, \quad \mathcal{T}^2 = -1, \\
\mathcal{C} &= iG_{102}\mathcal{K}, \quad \mathcal{C}^2 = -1, \\
\mathcal{S} &= G_{100}, \quad\quad \mathcal{S}^2 = +1.
\end{aligned} \tag{S78}$$

## B. A minimal model in Class C

Since a model in class C only preserves PHS, thus we consider this model consists of two copies of 4D QHIs in class A, i.e.,

$$\mathcal{H}_{\mathrm{C}}(k) = \begin{pmatrix} \mathcal{H}_{\mathrm{A}}^{+}(k) & 0 \\ 0 & \mathcal{H}_{\mathrm{A}}^{-}(k) \end{pmatrix}, \tag{S79}$$

where $\mathcal{H}_{\mathrm{A}}^{+}$ can simply obtained by introducing a perturbation to break TRS in $\mathcal{H}_{\mathrm{AII}}^{+}$ above. For example, we take

$$\mathcal{H}_{\mathrm{A}}^{+}(k) = \mathcal{H}_{\mathrm{AII}}^{+} + m_s G_{01}. \tag{S80}$$

For $\mathcal{C}^2 = -1$, we choose $\mathcal{C} = i\sigma_2 \otimes s\mathcal{K}$ where $s^* = s$ is a 4-dimensional Hermitian matrix. Similarly, we employ the relation and then obtain

$$\mathcal{C}\mathcal{H}_{\mathrm{C}}(k)\mathcal{C}^{-1} = -\mathcal{H}_{\mathrm{C}}(-k) \longrightarrow s\mathcal{H}_{\mathrm{A}}^{\mp *}(k)s^{-1} = -\mathcal{H}_{\mathrm{A}}^{\pm}(-k). \tag{S81}$$

Thus, for a given $\mathcal{H}_{\mathrm{A}}^{+}$, we have

$$\mathcal{H}_{\mathrm{A}}^{-}(k) = -s^{-1}\mathcal{H}_{\mathrm{A}}^{+ *}(-k)s. \tag{S82}$$

Assuming we pick $s = G_{33}$, then we obtain

$$\mathcal{H}_{\mathrm{A}}^{-}(k) = -d_x G_{31} + d_y G_{32} + d_z G_{33} + d_w G_{20} + d_m G_{10} + m_s G_{01}. \tag{S83}$$

We can check that this model hosts $C_2^- = -C_2^+$ when $m_s$ is small.

## S-VII.  REALIZATION PROPOSAL USING ULTRACOLD ATOMS

### A. Model Hamiltonian

Here we present the proposal for realizing the model Hamiltonian $\mathcal{H}_0$ of the main text. Since there are eight degrees of freedom in the model Hamiltonian, a candidate choice is the use of the alkaline-earth-metal atoms. It has a meta-stable excited-state manifold $|e\rangle$ besides the ground-state manifold $|g\rangle$ [11], from which we respectively select four hyperfine states as the pseudo-spins to construct the extra spaces. In particular, we specify the denotations to pseudo-spins as $\alpha = \uparrow\downarrow$ and $\beta = \pm$. The illustration of the setups is shown in FIG.S3. We prepare the eight states are coupled by three groups of optical fields. The Hamiltonian of the system is expressed in three parts,

$$H_0 = \hat{H}_0 + H_{\mathrm{so}}^{(\alpha)} + H_{\mathrm{so}}^{(\beta)}. \tag{S84}$$

The first part of Hamiltonian (S84) describes the local energy of the atoms,

$$\hat{H}_0 = \int d\mathbf{r} \sum_{\lambda,\lambda'} \sum_{\alpha,\beta} \psi_{\lambda,\alpha,\beta}^{\dagger}(\mathbf{r}) \left\{ \left[ -\frac{\nabla^2}{2m_{\mathrm{atom}}} + \hat{V}_{\mathrm{trap}}(\mathbf{r}) \right] \delta_{\lambda\lambda'} + \hat{V}[\sigma_3]_{\lambda\lambda'} \right\} \psi_{\lambda',\alpha,\beta}(\mathbf{r}). \tag{S85}$$

Here $\psi_{\lambda,\alpha,\beta}$ stands for the atomic operator for the $|\lambda = g, e\rangle$ manifold. The first term of Eq.(S85) describes the atomic kinetic energy. The second term is the lattice trap potential $\hat{V}_{\mathrm{trap}} = \sum_{\eta=x,y,z} V_L \sin^2(k_L \eta)$, where $V_L$ is the trap depth, $k_L = \pi/a$ and we set the lattice constant $a = 1$ and $\hbar = 1$ in the whole paper for simplicity of presentations. The last term describes the energy shifts with the magnitude $\hat{V}$ between different manifolds. We employ the tight-binding approximation. Eq.(S85) is then expressed as

$$\hat{H}_0 = \sum_j (-J\sigma_0 \otimes \sigma_0 \otimes \sigma_0 \psi_j^{\dagger}\psi_{j+1} + H.c.) + \hat{V}\sigma_3 \otimes \sigma_0 \otimes \sigma_0 \psi_j^{\dagger}\psi_j. \tag{S86}$$

Here $J$ is the nearest-neighbor hopping, $\psi_j = (e_{j\uparrow+}, e_{j\uparrow-}, e_{j\downarrow+}, e_{j\downarrow-}, g_{j\uparrow+}, g_{j\uparrow-}, g_{j\downarrow+}, g_{j\downarrow-})^T$ and $\lambda_j$ denotes the atomic annihilation operator on the $j$-th site.

The second part of Hamiltonian (S84) describes the coupling between the pseudo-spins $\alpha = \uparrow \downarrow$ [12]. It form is given as follows,

$$H_{so}^{(\alpha)} = \int d\mathbf{r} \sum_{\beta} M_1(\mathbf{r}) \left[ \psi_{e,\uparrow,\beta}^{\dagger}(\mathbf{r}) \psi_{g,\downarrow,\beta}(\mathbf{r}) + \psi_{e,\downarrow,\beta}^{\dagger}(\mathbf{r}) \psi_{g,\uparrow,\beta}(\mathbf{r}) \right] + H.c. \tag{S87}$$

We prepare the optical field as the following spatially modulated mode,

$$M_1(\mathbf{r}) = \mathcal{M}_1 \cos(k_L x) \sin(k_L y) \cos(k_L z) + \mathcal{M}_1' \cos(k_L x) \cos(k_L y) \cos(k_L z). \tag{S88}$$

We expand $\psi_{\lambda,\alpha,\beta}$ in terms of the Wannier wave function $W(\mathbf{r})$,

$$\psi_{\lambda,\alpha,\beta} = \sum_j W(\mathbf{r}) \lambda_j. \tag{S89}$$

In the tight-binding approximation. the on-site term introduced by $\mathcal{M}_1$ in Eq.(S88) vanishes because of the odd parity of $\sin(k_L y)$, therefore its nearest-neighbor term is dominant. Eq.(S87) is then written as

$$H_{so}^{(\alpha)} = \sum_{j,\beta} (-1)^{j_x+j_y+j_z} \frac{A_y}{2} (e_{j,\uparrow,\beta}^{\dagger} g_{j+e_y,\downarrow,\beta} - e_{j,\uparrow,\beta}^{\dagger} g_{j-e_y,\downarrow,\beta} + e_{j,\downarrow,\beta}^{\dagger} g_{j+e_y,\uparrow,\beta} - e_{j,\downarrow,\beta}^{\dagger} g_{j-e_y,\uparrow,\beta})$$
$$+ (-1)^{j_x+j_y+j_z} A_w (e_{j,\uparrow,\beta}^{\dagger} g_{j,\downarrow,\beta} + e_{j,\downarrow,\beta}^{\dagger} g_{j,\uparrow,\beta}) + H.c. \tag{S90}$$

where we have denoted

$$2\mathcal{M}_1 \int d\mathbf{r} \cos(k_L x) \sin(k_L y) \cos(k_L z) W^*(\mathbf{r}) W(\mathbf{r}+e_y) \equiv A_y \tag{S91}$$

and assigned $\mathcal{M}_1'$ with a cyclic parameter $k_w$ that is artificially controllable,

$$2\mathcal{M}_1' \int d\mathbf{r} \cos(k_L x) \cos(k_L y) \cos(k_L z) |W(\mathbf{r})|^2 \equiv A_w \sin(k_w). \tag{S92}$$

We remark that if the magnitudes of $A_y$ and $A_w$ are of the same order, the nearest-neighbor term introduced by $\mathcal{M}_1'$ in Eq.(S88) is negligible with respect to $A_y$.

The last part of Eq.(S84) describes the coupling between the pseudo-spins $\beta = \pm$,

$$H_{so}^{(\beta)} = \int d\mathbf{r} \sum_{\beta} M_2(\mathbf{r}) \left[ \psi_{e,\alpha,+}^{\dagger}(\mathbf{r}) \psi_{g,\alpha,-}(\mathbf{r}) + \psi_{e,\alpha,+}^{\dagger}(\mathbf{r}) \psi_{g,\alpha,-}(\mathbf{r}) - \psi_{e,\alpha,-}^{\dagger}(\mathbf{r}) \psi_{g,\alpha,+}(\mathbf{r}) - \psi_{e,\alpha,-}^{\dagger}(\mathbf{r}) \psi_{g,\alpha,+}(\mathbf{r}) \right] + H.c. \tag{S93}$$

Notice that in Eq.(S93) we have assumed a $\pi$ phase between the first two and last two coupling modes. This is attainable because they are generated by different optical fields, as shown in FIG.S3(a). We prepare the optical field mode as,

$$M_2(\mathbf{r}) = i\mathcal{M}_2 \sin(k_L x) \cos(k_L y) \cos(k_L z) + \mathcal{M}_2 \cos(k_L x) \cos(k_L y) \sin(k_L z), \tag{S94}$$

Likewise as performed in obtaining Eq.(S90), we obtain the tight-binding form,

$$H_{so}^{(\beta)} = \sum_{j,\alpha} (-1)^{j_x+j_y+j_z} \frac{iA_x}{2} (e_{j,\alpha,+}^{\dagger} g_{j+e_x,\alpha,-} - e_{j,\alpha,+}^{\dagger} g_{j-e_x,\alpha,-} - e_{j,\alpha,-}^{\dagger} g_{j+e_x,\alpha,+} + e_{j,\alpha,-}^{\dagger} g_{j-e_x,\alpha,+})$$
$$+ (-1)^{j_x+j_y+j_z} \frac{A_z}{2} (e_{j,\alpha,+}^{\dagger} g_{j+e_z,\alpha,-} - e_{j,\alpha,+}^{\dagger} g_{j-e_z,\alpha,-} - e_{j,\alpha,-}^{\dagger} g_{j+e_z,\alpha,+} + e_{j,\alpha,-}^{\dagger} g_{j-e_z,\alpha,+}) + H.c. \tag{S95}$$

where we have denoted

$$2\mathcal{M}_2 \int d\mathbf{r} \sin(k_L x) \cos(k_L y) \cos(k_L z) W^*(\mathbf{r}) W(\mathbf{r}+de_x) \equiv A_x, \tag{S96}$$

$$2\mathcal{M}_2 \int d\mathbf{r} \cos(k_L x) \cos(k_L y) \sin(k_L z) W^*(\mathbf{r}) W(\mathbf{r}+de_z) \equiv A_z. \tag{S97}$$

As shown in FIG.S3(a), the couplings between the manifolds $|g\rangle$ and $|e\rangle$ are specified into three groups. We prepare three configurations of two-photon transitions, which are shown in FIG.S3(b)-(c). By taking the advantages of the selection rules during the atomic transitions, we rearrange the pseudo-spin representations for the states of each manifold. This makes it possible to separately engineer the couplings via different transitions.

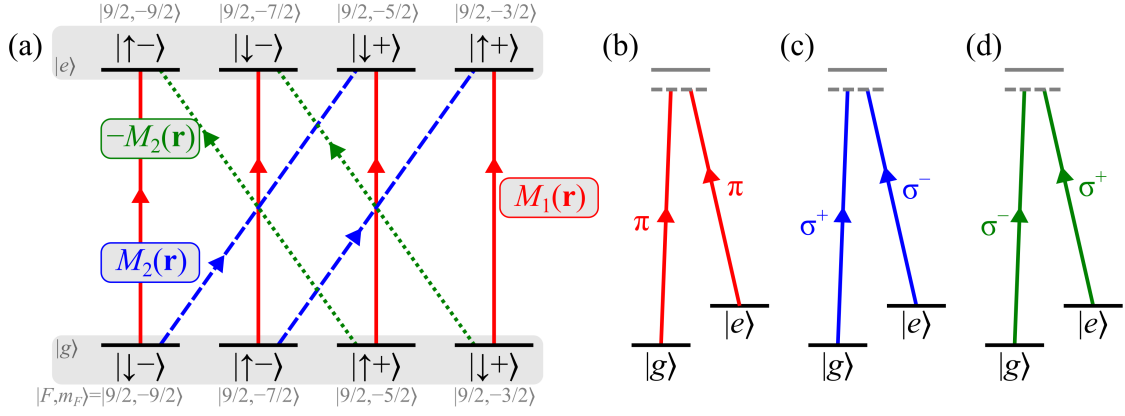

FIG. S3. (a) Illustration of the realization setups. We use $^{87}$Sr as the example [13] and choose its state manifolds $^1S_0$ and $^3P_0$ as $|g\rangle$ and $|e\rangle$. The red-solid, blue-dashed, and green-dotted lines show the coupling with field modes of $M_1(\mathbf{r})$, $M_2(\mathbf{r})$, and $-M_2(\mathbf{r})$, respectively. They are processed between the atomic hyperfine states $|F, m_F\rangle$ of different manifolds. (b) Setups of the two-photon transition for generating the field mode $M_1(\mathbf{r})$. It is constructed by two $\pi$-polarized fields. Each state of $|g\rangle$ transits to another of $|e\rangle$ via an inter-mediate state. There is no changes of the angular momenta during the transition. (c) Setups for $M_2(\mathbf{r})$. It is constructed by two fields with opposite $\sigma$ polarizations. It increases the angular momenta from $|g\rangle$ to $|e\rangle$. (d) Setups for $-M_2(\mathbf{r})$. It decreases the angular momenta from $|g\rangle$ to $|e\rangle$.

## B. Operator representations

We make the following operator representation solely for the $|e\rangle$ manifold,

$$e_{j,\alpha,\beta} \to (-1)^{j_x+j_y+j_z} e_{j,\alpha,\beta} \,. \tag{S98}$$

It does not violate the anti-commutation relation of the original atomic operators. Under the representation (S98) and transformed to the momentum space, Eq.(S86) becomes

$$\hat{H}_0(\mathbf{k}) = \{2J[\cos(k_x)+\cos(k_y)+\cos(k_z)] + \hat{V}\}\sigma_3 \otimes \sigma_0 \otimes \sigma_0 \,. \tag{S99}$$

Eq.(S87) becomes

$$H_{\text{so}}^{(\alpha)}(\mathbf{k}) = A_y \sin(k_y)\sigma_1 \otimes \sigma_2 \otimes \sigma_3 + A_w \sin(k_w)\sigma_1 \otimes \sigma_1 \otimes \sigma_3 \,. \tag{S100}$$

Eq.(S93) becomes

$$H_{\text{so}}^{(\beta)}(\mathbf{k}) = A_x \sin(k_x)\sigma_1 \otimes \sigma_0 \otimes \sigma_1 + A_z \sin(k_z)\sigma_1 \otimes \sigma_0 \otimes \sigma_2 \,. \tag{S101}$$

We prepare

$$A_x = A_y = A_z = A_w = -2J = 1 \,, \tag{S102}$$

and the cyclic dependence of $\hat{V}$ on $k_w$,

$$\hat{V} = m - \cos(k_w) \,. \tag{S103}$$

The parameter $k_w$ can be recognized as the momentum of the synthetic dimension. After making $\sigma_2 \leftrightarrow \sigma_3$ in the last $\sigma$-matrix space, we obtain the effective Hamiltonian by combining Eqs.(S99)-(S101),

$$\mathcal{H}_0(\mathbf{k}) = \sum_{i=x,y,z,w} d_i(\mathbf{k})\Gamma_i + d_m(\mathbf{k})\Gamma_0 \tag{S104}$$

with

$$
\begin{cases}
d_i(\mathbf{k}) = \sin(k_i) \\
d_m(\mathbf{k}) = m - \sum_i \cos(k_i)
\end{cases}
,\qquad
\begin{cases}
\Gamma_1 = \sigma_1 \otimes \sigma_0 \otimes \sigma_1 \\
\Gamma_2 = \sigma_1 \otimes \sigma_2 \otimes \sigma_2 \\
\Gamma_3 = \sigma_1 \otimes \sigma_0 \otimes \sigma_3 \\
\Gamma_4 = \sigma_1 \otimes \sigma_1 \otimes \sigma_2 \\
\Gamma_0 = \sigma_3 \otimes \sigma_0 \otimes \sigma_0
\end{cases}
. \tag{S105}
$$




\begin{thebibliography}{99}
\bibitem{note0} The matrices are given by
		$\Gamma_1=\sigma_1\otimes\sigma_0\otimes\sigma_1$, $\Gamma_2=\sigma_1\otimes\sigma_2\otimes\sigma_2$,
		$\Gamma_3=\sigma_1\otimes\sigma_0\otimes\sigma_3$, $\Gamma_4=\sigma_1\otimes\sigma_1\otimes\sigma_2$, and
		$\Gamma_0=\sigma_3\otimes\sigma_0\otimes\sigma_0$.
\bibitem{note3} The block diagonal Hamiltonian reads
		$u\mathcal{H}_{RD}u^{-1}=\mathcal{H}_W^+\oplus \mathcal{H}_W^-$,
		with $u=\exp(-i\frac{\pi}{4}G_{10})\exp(i\frac{\pi}{4}G_{12})$.
		Here
		$\mathcal{H}_{W}^{\pm}=k_x\sigma_1\pm k_y\sigma_2- k_z\sigma_3=\sum_{i,j}v^{\pm}_{ij}k_i\sigma_j$.
		The chirality can be verified as $\eta_{\pm}=\text{sign}(\det v^{\pm}_{ij})=\mp1$.
\bibitem{note6} In the region $0<m<2[-2<m<0]$, there are three real Dirac points (or three pairs of Weyl points) that survive on the 3D boundary near three high symmetry points $(\pi,0,0)[(\pi,\pi,0)]$, $(0,\pi,0)[(\pi,0,\pi)]$, $(0,0,\pi)[(0,\pi,\pi)]$, respectively. In the region $-4<m<-2$, a real Dirac point (or a pair of Weyl points) survives on the 3D boundary near $(\pi,\pi,\pi)$.
\end{thebibliography}
\end{document}